\documentclass[lineno]{jfm}

\usepackage{xcolor}
\usepackage{graphicx}
\usepackage{amsmath}
\usepackage{amssymb}
\usepackage{pifont}
\usepackage{txfonts}
\usepackage{psfrag}
\usepackage[round]{natbib}
\usepackage[T1]{fontenc}
\usepackage{lmodern}
\usepackage{tikz}
\usetikzlibrary{shapes.geometric,shapes.symbols}
\usepackage{scalerel}
\usepackage{upgreek}
\usepackage{booktabs}
\usepackage{verbatim}
\usepackage{framed}
\usepackage{fancyvrb}
\usepackage{varwidth}
\usepackage{bbding}
\usepackage{subfigure}
\usepackage{multirow}
\usepackage[normalem]{ulem}
\usepackage{multicol}
\usepackage{breqn}
\usepackage{soul}
\usepackage{dirtytalk}
\usepackage{upquote}
\usepackage{authblk}
\usepackage{array}

\definecolor{OliveGreen}{RGB}{0,127,0}
\definecolor{Golden}{RGB}{222,127,0}
\definecolor{Teal}{RGB}{0,127,127}

\setstcolor{blue}

\newcolumntype{P}[1]{>{\centering\arraybackslash}p{#1}}

\usepackage{hyperref}
\hypersetup{
    colorlinks = true,
    urlcolor   = blue,
    citecolor  = black,
}

\newcommand{\RomanNumeralCaps}[1]
\linenumbers

\title{Evidence that superstructures comprise of self-similar coherent motions in high $Re_{\tau}$ boundary layers}
\author{Rahul Deshpande\aff{1}
  \corresp{\email{raadeshpande@gmail.com}},
	Charitha M. de Silva\aff{2},
 and Ivan Marusic\aff{1}}

\affiliation{\aff{1}Department of Mechanical Engineering, University of Melbourne, Parkville, VIC 3010, Australia
\aff{2}School of Mechanical and Manufacturing Engineering, University of New South Wales, Sydney, NSW 2052, Australia}

\begin{document}
\maketitle

\begin{abstract}

We present experimental evidence that the superstructures in turbulent boundary layers comprise of smaller, geometrically self-similar coherent motions. 
The evidence comes from identifying and analyzing instantaneous superstructures from large-scale particle image velocimetry datasets acquired at high Reynolds numbers, capable of capturing streamwise elongated motions extending up to 12 times the boundary layer thickness. 
Given the challenge in identifying the constituent motions of the superstructures based on streamwise velocity signatures, a new approach is adopted that analyzes the wall-normal velocity fluctuations within these very long motions, which reveals the constituent motions unambiguously. 
The conditional streamwise energy spectra of the wall-normal fluctuations, corresponding exclusively to the superstructure region, are found to exhibit the well-known distance-from-the-wall scaling in the intermediate scale range. 
Similar characteristics are also exhibited by the Reynolds shear stress co-spectra estimated for the superstructure region, suggesting that geometrically self-similar motions are the constituent motions of these very-large-scale structures. 
Investigation of the spatial organization of the wall-normal momentum-carrying eddies, within the superstructures, also lends empirical support to the concatenation hypothesis for the formation of these structures.
The association between the superstructures and self-similar motions is reaffirmed on comparing the vertical coherence of the Reynolds-shear-stress carrying motions, by computing conditionally-averaged two-point correlations, which are found to match with the mean correlations. 
The mean vertical coherence of these motions, investigated for the log-region across three decades of Reynolds numbers, exhibits a unique distance-from-the-wall scaling invariant with Reynolds number. 
The findings support the prospect for modelling these dynamically significant motions via data-driven coherent structure-based models.

\end{abstract}


\section{Introduction and motivation}
\label{intro}

Over the past two decades, the study of high Reynolds number ($Re_{\tau}$ $\gtrsim$ $\mathcal{O}$(10$^4$)) wall-bounded flows has become synonymous with very-large-scale motions (VLSMs), also known as `superstructures', which play a predominant role in the dynamics and spatial organization of wall turbulence.
Here, $Re_{\tau}$ $=$ ${\delta}{U_{\tau}}/{\nu}$, where $\delta$ is the boundary layer thickness, $\nu$ is the kinematic viscosity and $U_{{\tau}}$ is the skin-friction velocity, with the latter two used to normalize the statistics in viscous units (indicated by superscript `$+$').
The superstructures can extend beyond 20$\delta$ in the streamwise direction \citep{kim1999,hutchins2007} and also exhibit `meandering' when viewed on a wall-parallel plane \citep{charitha2015}, particularly in the logarithmic region of the flow. Such a large spatial footprint permits these motions to carry significant proportions of the total turbulent kinetic energy and the Reynolds shear stresses of the flow \citep{liu2001,guala2006,balakumar2007}. 
Given that the shear stress is responsible for the wall-normal momentum transfer, this suggests that the VLSMs/superstructures also contribute significantly to the high $Re_{\tau}$ turbulent skin-friction drag \citep{deck2014}.
Hence, an improved understanding of the origin of these VLSMs/superstructures, towards which this study is directed, stands to advance our knowledge in both a fundamental and an applied perspective.

\begin{figure}
   \captionsetup{width=1.0\linewidth}
  \centerline{\includegraphics[width=1.0\textwidth]{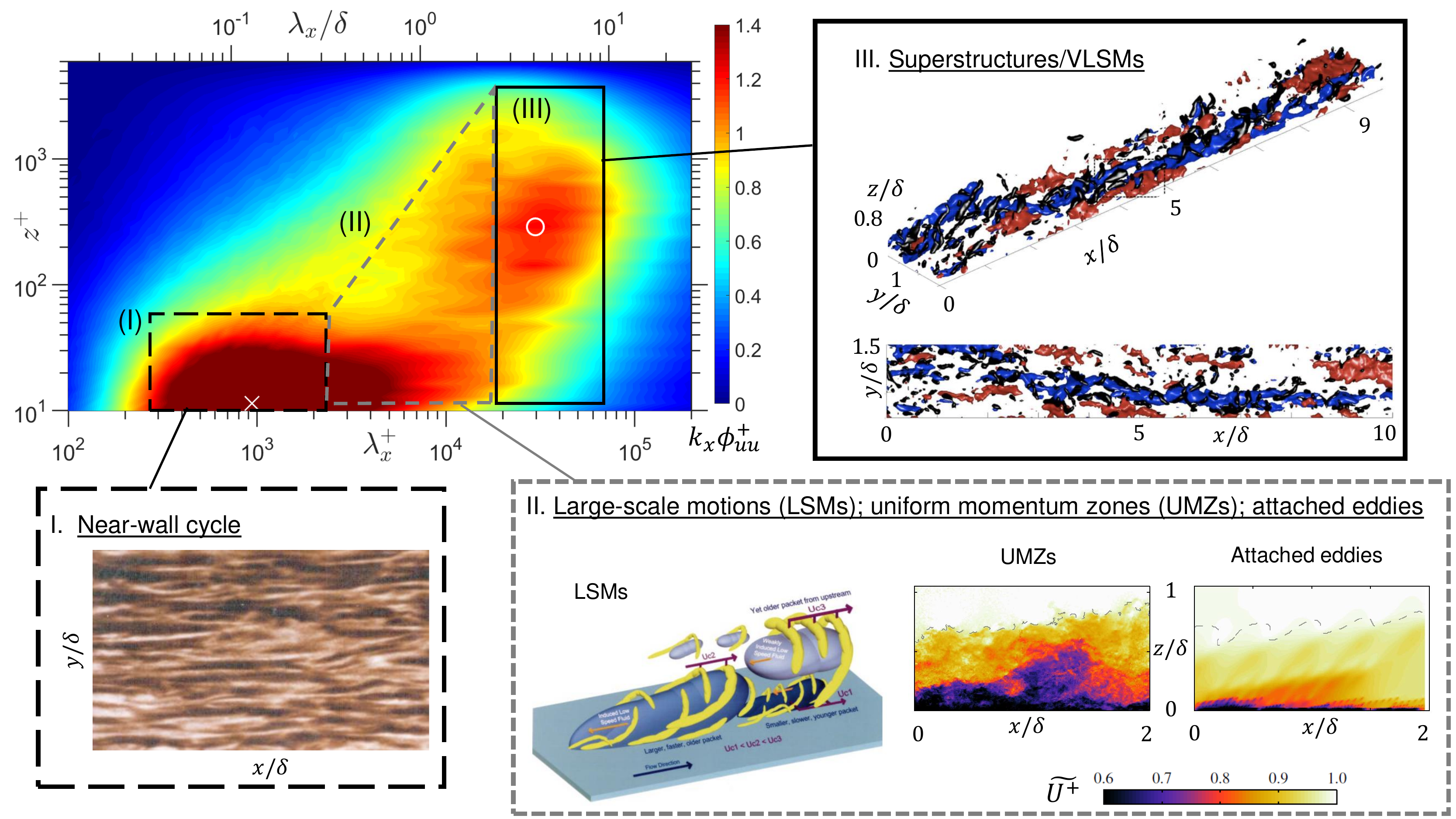}}
  \caption{Premultiplied spectra of the streamwise velocity (${k_{x}}{{\phi}^{+}_{uu}}$) plotted against viscous-scaled wavelength (${\lambda}^{+}_{x}$) and distance from the wall ($z^+$) for a turbulent boundary layer at $Re_{\tau}$ $\approx$ 7300 \citep{hutchins2007}. $\times$ and $\bigcirc$ marked in the plot correspond to the `inner' and `outer' peaks of the $u$-spectrogram noted previously in the literature. 
	Regions (I), (II) and (III) are used to indicate spectral signatures of various coherent motions observed in the literature. 
	Region (I) corresponds to the near-wall cycle captured via flow visualization by Prof. S. J. Kline (photo shared by Prof. D. Coles). Region (II) corresponds to the LSMs (conceptual sketch by \citealp{adrian2000}), UMZs (particle image velocimetry (PIV) by \citealp{charitha2016}) and attached eddies (attached eddy simulations by \citealp{charitha2016}). 
	Region (III) corresponds to the VLSMs/superstructures, visualized via time resolved PIV by \citet{dennis2011Part2}.}
\label{fig1}
\end{figure}

\citet{hutchins2007} used the terminology `superstructures' when referring to the spectrogram of the streamwise velocity fluctuations ($u$) from a high $Re_{\tau}$ boundary layer, as shown in figure \ref{fig1}. 
The spectrogram presents  the premultiplied $u$-energy spectra as a function of the viscous-scaled streamwise wavelengths (${\lambda}^{+}_{x}$ $=$ ${{\lambda}_{x}}{U_{\tau}}/{\nu}$) and wall-normal distance ($z^+$ $=$ $z{U_{\tau}}/{\nu}$), with ${\lambda}_{x}$ = 2$\pi$/${k_x}$, where $k_x$ is the streamwise wavenumber.
The high $Re_{\tau}$ $u$-spectrogram is seen to have two prominent peaks. 
One is located in the inner-region synonymous with the well-documented near-wall cycle \citep{kline1967}, consisting of high and low-speed viscous-scaled streaks (${\lambda}^{+}_{x}$ $\approx$ 1000), which are responsible for intense local production of turbulent kinetic energy.
The second peak is in the outer region of the flow (typically in the logarithmic/inertial region), and corresponds to the superstructures, which have a spectral signature at very long wavelengths (${\lambda}_{x}$ $\sim$ 6${\delta}$) and also extend down to the wall \citep{hutchins2007}.
It is worth noting here that this second peak is only visible for $Re_{\tau}$ $\gtrsim$ 2000, owing to the insufficient separation of scales and weaker energy of the superstructures at lower $Re_{\tau}$ \citep{hutchins2007}.
Between the inner- and outer-peaks, a nominal plateau is seen in the spectrogram which corresponds to the distance-from-the-wall ($z$)-scaled eddies coexisting in the log-region; these eddies make up the increased range of scales with increasing $Re_{\tau}$.
In the literature, these intermediate scaled eddies have been described by various structures or motions, including the large-scale motions (LSMs; \citealp{kim1999}, \citealp{adrian2000}), uniform momentum zones (UMZs; \citealp{meinhart1995}, \citealp{charitha2016}), attached eddies \citep{baars2017,marusic2019,hu2020,deshpande2021} and so forth.
In the remainder of this section, for simplicity, we will refer to these motions as LSMs. 
It should also be noted that the terminology `VLSMs' and `superstructures' have been conventionally associated with the very-large-scale motions in internal \citep{kim1999} and external wall-bounded flows \citep{hutchins2007}, respectively.
Considering this study focuses solely on zero-pressure gradient turbulent boundary layers, we henceforth refer to either of these structures simply as superstructures.

To date, several studies have investigated the probable mechanisms responsible for the formation of superstructures, with two theories hypothesized most often: (i) the formation of superstructures via concatenation of the LSMs \citep{kim1999,adrian2000,lee2011,dennis2011Part2}, or (ii) the emergence of superstructures due to a linear instability mechanism \citep{del2006linear,mckeon2010,hwang2010}.
The present study does not focus on comparing and contrasting the likelihood of one mechanism over the other.
Rather, it builds upon recent compelling evidence in support of the concatenation mechanism \citep{wu2012,baltzer2013,lee2014,lee2019}, to investigate the characteristics of the constituent motions forming the superstructures.
The formation of superstructures via streamwise concatenation of the relatively smaller motions has been confirmed by several studies conducted across all canonical wall-bounded flows (turbulent boundary layers, channels, pipes), through: (i) investigation of the time evolution of instantaneous flow fields \citep{lee2011,dennis2011Part2,wu2012,lee2019}, (ii) statistical analysis of the superstructure formation frequency/population density \citep{lee2014} and (iii) spatial correlations of the low-pass filtered velocity fields \citep{baltzer2013,lee2019}.
In comparison, few studies have presented similar statistical arguments in favour of the linear instability mechanism. 
For instance, \citet{bailey2008} supported the linear instability argument by noting different spanwise widths of the superstructures and LSMs in the inertial region of a turbulent pipe flow.
Their estimates, however, were limited to two-point velocity correlations reconstructed in a particular wall-parallel plane, which cannot be uniquely associated with the LSMs responsible for the superstructure formation \citep{deshpande2020}.
Considering that superstructures extend down from the log-region to the wall, \citet{deshpande2021model} reconstructed two-point velocity correlations across two wall-parallel planes located in the near-wall and the log-region.
These statistics, which are purely representative of the large `wall-coherent' motions, revealed similar spanwise extents of the coexisting superstructures and LSMs for all canonical wall flows, thereby favouring the concatenation argument.

Despite substantial support for the concatenation argument, several unanswered questions are still associated with this mechanism.
For instance, there is no universal agreement on what facilitates the streamwise concatenation of LSMs to form superstructures.
While few studies have associated this with the spanwise alternate positioning of low and high momentum LSMs \citep{lee2014}, others have conjectured the role played by secondary roll cells \citep{baltzer2013,lee2019} in favourably organizing the relatively smaller motions.
Progress in this regard has been hindered by the lack of understanding of the constituent motions forming the superstructures; 
for instance, are superstructures purely composed of the inertial $\delta$-scaled motions corresponding to the extreme right end of region II in figure \ref{fig1}?
Or do they also comprise of the geometrically self-similar, i.e. $z$-scaled hierarchy of eddies encompassing the entirety of region II? 
The present study aims to answer these questions by analyzing the characteristics of the constituent motions.

In the past, clarifying such information on the constituent motions has not been possible due to the low to moderate $Re_{\tau}$ ($\lesssim$ 2000) of the experiments/simulations analyzing the concatenation argument, which severely constricts the extent of region (II) in figure \ref{fig1}.
This prevents an unambiguous delineation between the $\delta$-scaled and $z$-scaled inertial motions coexisting in region II.
Further, the statistical signature of the superstructures is also very weak at these $Re_{\tau}$ \citep{hutchins2007}, making it challenging to identify and isolate them from the other motions in the flow.
However, increased access to high $Re_{\tau}$ data over the past decade has substantially increased our knowledge of these inertial eddies coexisting in the log and outer regions \citep{marusic2015,baidya2017,deshpande2021}. 
This has also led to growing acceptance of the existence of the geometrically self-similar attached eddy hierarchy in the inertial region \citep{charitha2016,baars2017,hwang2018,hu2020,deshpande2020,deshpande2021}, which can be modelled conceptually \citep{marusic2019}.
These advancements make it compelling to investigate whether these self-similar inertial motions are associated with the formation of superstructures, a conjecture that has previously shown promising results when tested for low $Re_{\tau}$ channel flows \citep{lozano2012}, and when implemented in coherent structure-based models \citep{deshpande2021model}.
If this conjecture is proven true, then the preferred streamwise alignment of this energy-containing hierarchy of motions (to form superstructures) would have implications on Townsend's attached eddy hypothesis, which otherwise assumes a random distribution of attached eddies in the flow field \citep{townsend1976,marusic2019}.
The investigation can also help answer the long-standing contradiction \citep{guala2006,balakumar2007,wu2012} between: (i) the attached eddy hypothesis, which classifies turbulent superstructures to be `inactive' \citep{deshpande2021}, and (ii) instantaneous flow field observations, per which these streamwise elongated motions carry significant Reynolds shear stresses (and hence behave as `active' motions).

\begin{figure}
   \captionsetup{width=1.0\linewidth}
  \centerline{\includegraphics[width=1.0\textwidth]{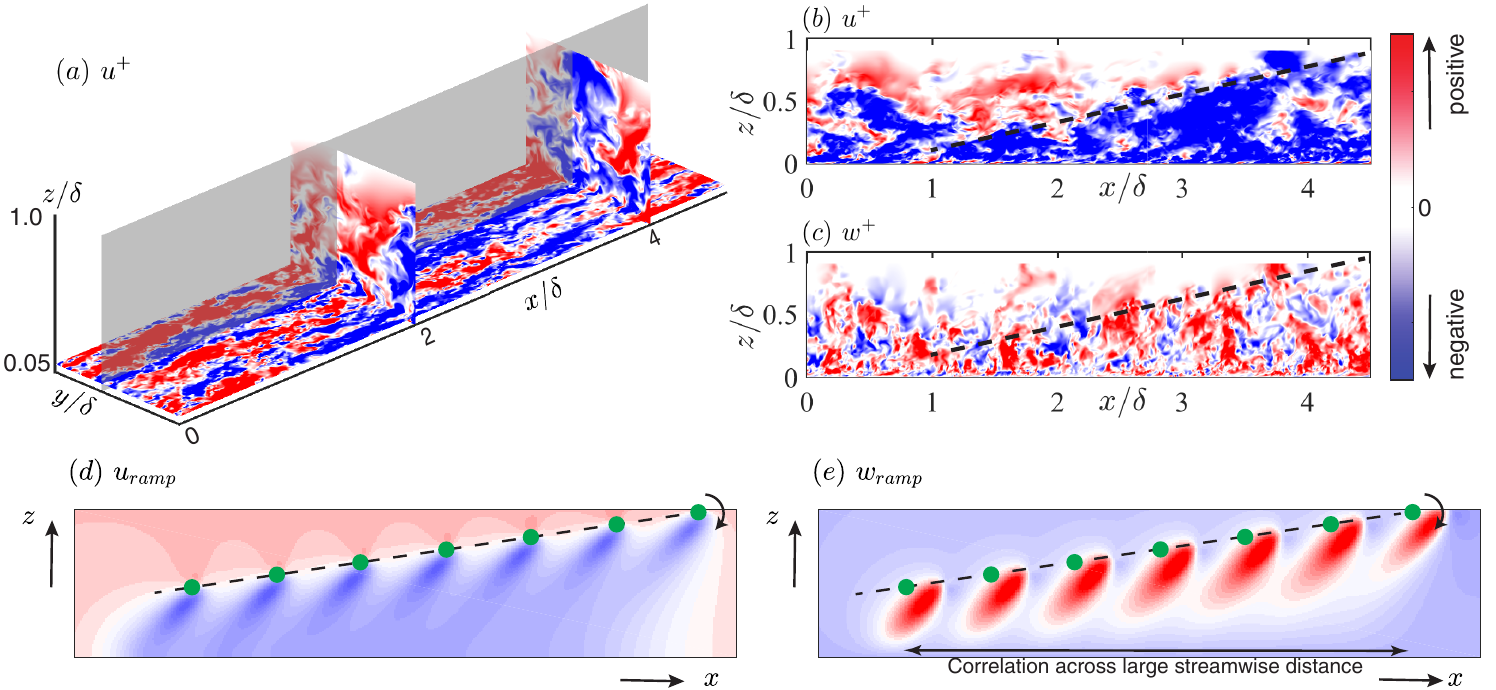}}
  \caption{Colour contours of the instantaneous (a,b) streamwise, $u$ and (c) wall-normal velocity fluctuations, $w$ in a boundary layer at $Re_{\tau}$ $\approx$ 2000.
	This data has been extracted from a particular 3-D time block of the publicly available DNS dataset of \citet{sillero2013}.
	In (a), $u$ is plotted on a wall-parallel plane at $z$ $\approx$ 0.05$\delta$, as well as on the cross-planes at $x$ $\approx$ 2$\delta$ and 4$\delta$.
 (b) and (c) respectively plot the $u$ and $w$ fluctuations in the streamwise wall-normal plane shaded in grey in (a).
	Dashed black line in (b,c) traces the top part of a long $-u$ ramp type structure.
	(d,e) respectively plot an idealized distribution of $u$ and $w$ flow field induced by multiple prograde vortices (in green) positioned along the ramp \citep{adrian2000,charitha2016}.}
\label{fig2}
\end{figure}

To this end, the present study investigates the geometric scalings exhibited by the constituent motions of the superstructures.
Experimental data is employed from a moderate to high $Re_{\tau}$ turbulent boundary layer (2500 $\lesssim$ $Re_{\tau}$ $\lesssim$ 7500), which is an order of magnitude higher than the simulation studies reported previously, to ensure coexistence of a broad range of inertial scales (region II).
The dataset comprises of sufficiently resolved large-scale velocity fluctuations acquired in a physically thick boundary layer via unique, large field-of-view (LFOV) particle image velocimetry (PIV), capturing instantaneous flow fields with an extent of 12$\delta$ in the streamwise direction ($x$).
In contrast to most studies to date, which have investigated the superstructures by analyzing the large-scale $u$-fluctuations, here we adopt a unique strategy to investigate the wall-normal ($w$) velocity fluctuations within the superstructure region.
This is because deciphering smaller constituent $u$-motions from within a larger $u$-motion can be inconclusive, as can be noted from a sample DNS flow field shown in figures \ref{fig2}(a,b).
On the other hand, the $w$-fluctuations can bring out the individual constituent motions more distinctly, which is evident from figure \ref{fig2}(c) and will be analyzed here by computing conditional statistics.
It can be noted from figures \ref{fig2}(a-c) that the individual $w$-eddies within the region associated with a long $u$-motion are much smaller in streamwise extent (than $u$), and exhibit sort of a clustered/packed organization plausibly leading to the appearance of a $u$-superstructure.
This scenario is recreated in figure \ref{fig2}(d,e), using an idealized distribution of prograde vortices, which suggests the possibility of strong $u$- as well as $w$-correlations extending across large streamwise separations.
Such a flow organization, which adds further credibility to the streamwise concatenation hypothesis, will be investigated here via conditional statistics from high $Re_{\tau}$ data.
It is important to note that in the present study, any reference to concatenation henceforth refers to the spatial organization of constituent motions over extended streamwise distances, such as in figures \ref{fig2}(d,e).
Given the experimental limitations, the study cannot directly comment on the dynamics/mechanism behind how this spatial organization comes into existence.
Also, the terminology `attached eddies' is used here to refer to any eddies/motions scaling with their distance from the wall, and hence is not limited to the eddies physically extending to the wall.

\begin{table}
   \captionsetup{width=1.0\linewidth}
\centering
\begin{center}
\begin{tabular}{@{\extracolsep{1pt}} P{1.7cm} P{1.5cm} P{1.1cm} P{1.1cm} P{0.7cm} P{0.7cm} P{1.5cm} P{3.8cm}}
{Measurement} & {Facility} & {$Re_{\tau}$} & {${\nu}/{U_{\tau}}$} & ${\Delta}{{x}^{+}}$ & ${\Delta}{{z}^{+}}$ & FOV & Reference  \\
  &  &  & (in $\mu$m) &  &  & ($x$ $\times$ $z$) & \\
LFOV PIV & HRNBLWT & 2500 & 42 & 26 & 26 & 12$\delta$ $\times$ 1.2$\delta$ & \citet{charitha2015,charitha2020} \\
LFOV PIV & HRNBLWT & 5000 & 22 & 52 & 52 & 12$\delta$ $\times$ 1.2$\delta$ & \citet{charitha2015,charitha2020} \\
LFOV PIV & HRNBLWT & 7500 & 15 & 75 & 75 & 12$\delta$ $\times$ 1.2$\delta$ & \citet{charitha2015,charitha2020} \\
PIV & HRNBLWT & 14500 & 24 & 37 & 37 & 2$\delta$ $\times$ 0.4$\delta$ & \citet{charitha2014} \\
Sonics & SLTEST & $\mathcal{O}$(10$^6$) & 78 & 1000 & 90 & -- & \citet{hutchins2012} \\
\hline
\end{tabular}
\caption{Table summarizing details of datasets comprising synchronized measurements of $u$- and $w$-fluctuations at various wall-normal locations. 
$Re_{\tau}$ for the various PIV datasets is based on $\delta$ estimated at the centre of the flow field (figure \ref{fig3}a).
Terminology has been defined in $\S$\ref{exp}. 
${\Delta}{x^+}$ and ${\Delta}{z^+}$ indicate viscous-scaled spatial resolution along $x$ and $z$ directions, respectively.} 
\label{tab1}
\end{center}
\end{table}

\section{Experimental datasets and methodology}
\label{exp}

\subsection{Description of the experimental datasets}
\label{data}

Five multipoint datasets are used from previously published high $Re_{\tau}$ experiments (table \ref{tab1}).
Four of these are acquired via two-dimensional (2-D) two-component PIV in the Melbourne wind tunnel (HRNBLWT; \citealp{marusic2015}) and span the $Re_{\tau}$ range $\sim$ 2500--14500.
The test section of this wind tunnel has a cross-section of 0.92\;m $\times$ 1.89\;m, and has a large streamwise development length of $\sim$27\;m, with maximum possible free-stream speeds ($U_{\infty}$) of up to 45\;ms$^{-1}$.
Such a large-scale facility permits the generation of a sufficiently high $Re_{\tau}$ canonical boundary layer flow facilitated by substantial increment in its boundary layer thickness, along its long streamwise fetch.
This capability is leveraged in the four PIV datasets employed in the present study, which will be described next.

\begin{figure}
   \captionsetup{width=1.0\linewidth}
  \centerline{\includegraphics[width=1.0\textwidth]{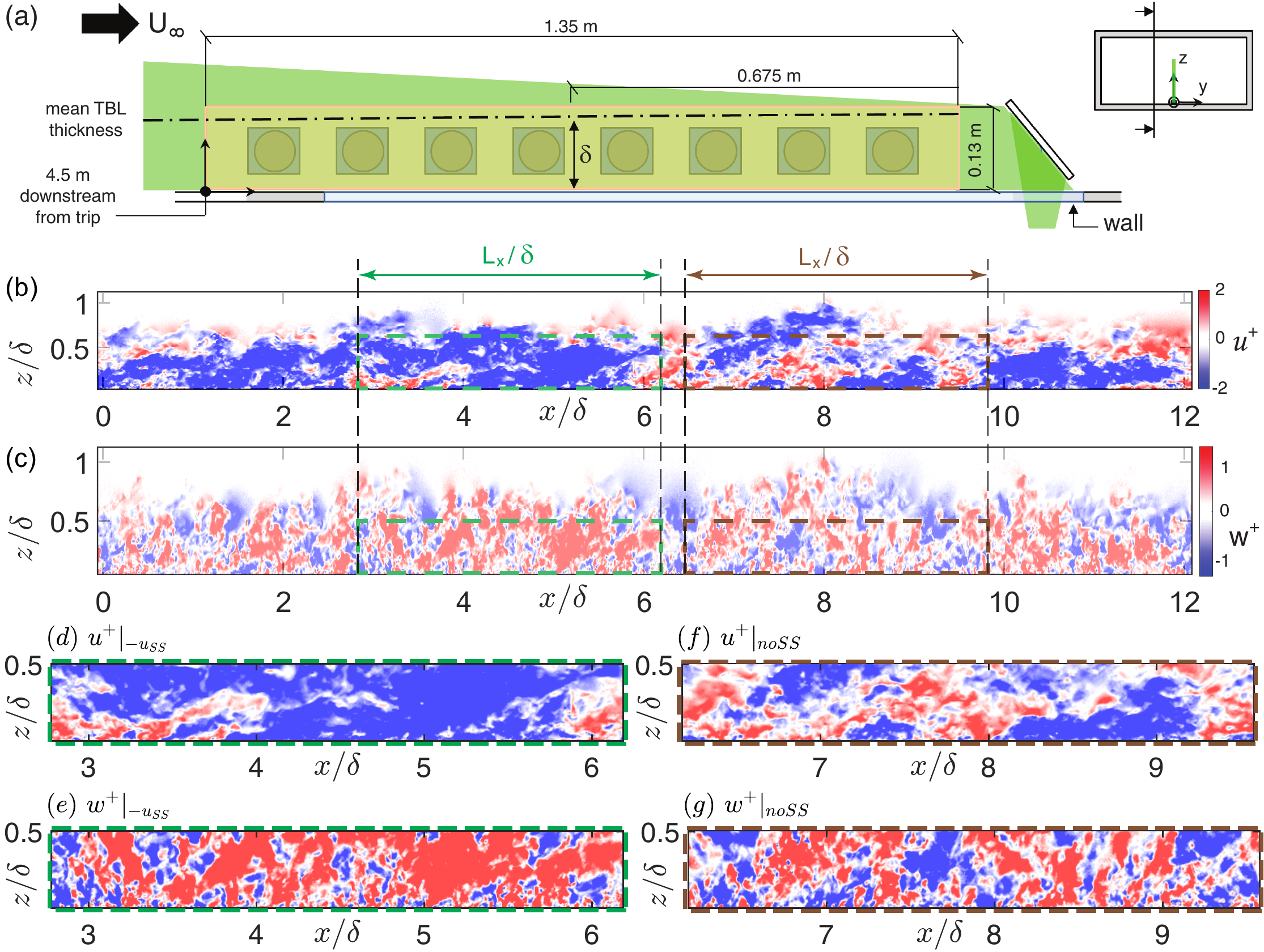}}
  \caption{Schematic of the experimental setup used to conduct LFOV PIV experiments in the streamwise wall-normal plane ($x$--$z$) in the HRNBLWT. 
Green shading indicates flow illuminated by the laser while the orange shading indicates the flow field cumulatively captured by the PIV cameras (shown in the background). 
Dash-dotted black line represents the streamwise evolution of the boundary layer thickness, with $\delta$ defined at the centre of the full flow field.
(b,c) Instantaneous (b) $u^+$ and (c) $w^+$-fluctuations from the LFOV PIV dataset at $Re_{\tau}$ $\approx$ 2500.
The {\color{OliveGreen}dashed green} box in (b,c) identifies a low-momentum turbulent superstructure ($-{u_{ss}}$) of length $L_{x}$ based on the superstructure extraction algorithm described in $\S$\ref{extract}.
(d,e) show an expanded view of the $u$- and $w$-fluctuations within $-{u_{ss}}$, as identified in (b,c), respectively.
Alternatively, the {\color{brown}dashed brown} box in (b,c) represents flow field of the same length$\times$height as the dashed green box, but not associated with a turbulent superstructure (${noSS}$).
(f,g) show an expanded view of the $u$- and $w$-fluctuations within the ${noSS}$ region identified in (b,c), respectively.}	
\label{fig3}
\end{figure}

Three of the PIV datasets comprise snapshots of very large streamwise wall-normal flow fields of a turbulent boundary layer ($x$ $\times$ $z$ $\sim$ 12$\delta$ $\times$ 1.2$\delta$), and are thus henceforth referred to as the large field-of-view (LFOV) PIV datasets \citep{charitha2015,charitha2020}. 
To the best of the authors' knowledge, this is the only published lab-based dataset giving access to sufficient LFOV instantaneous flow fields at $Re_{\tau}$ $\gtrsim$ 5000 (to achieve statistical convergence), thereby making the analysis presented in this paper unique as well as ideally-suited for investigating turbulent superstructures.
The LFOV is made possible by stitching the imaged flow fields from eight high-resolution 14 bit PCO 4000 PIV cameras, each with a sensor resolution of 4008 $\times$ 2672 pixels.
Figure \ref{fig3}(a) shows a schematic of the experimental setup for the LFOV PIV, where the region shaded in orange indicates the individual FOVs combined from the eight cameras.
These measurements were conducted at the upstream end of the test section, with the LFOV starting at $x$ $\approx$ 4.5\;m from the start of the test section. 
The experiments were conducted at three free-stream speeds ($U_{\infty}$ $\approx$ 10, 20 and 30\;m{s$^{-1}$}), which led to a corresponding variation in $Re_{\tau}$ of 2500, 5000 and 7500, respectively.
Here, $U_{\tau}$ and $\delta$ used to estimate the flow $Re_{\tau}$, were computed at the middle of the LFOV, using the method outlined in \citet{chauhan2009}.
The boundary layer thickness is nominally $\delta$ $\approx$ 0.11\;m for all three $Re_{\tau}$ cases.

Considering the focus of the experiment was on a LFOV, a homogeneous seeding density was ensured across the entire test section of the tunnel for these measurements, and the particles were illuminated by a Big Sky Nd-YAG double pulse laser ($\sim$1 \.mm thickness), delivering 120\;mJ/pulse.
The last optical mirror to direct this laser sheet was tactically placed within the test section (figure \ref{fig3}a), for ensuring adequate laser illumination levels across the LFOV.
This optic arrangement, however, was sufficiently downstream of the PIV flow field and introduced no adverse effects (such as blockage, etc) on the measurement \citep{charitha2015}. 
Figures \ref{fig3}(b,c) gives an example of the viscous-scaled $u$- and $w$-fluctuations estimated from the LFOV PIV experiment at $Re_{\tau}$ $\approx$ 2500, which successfully captures a turbulent superstructure (of length $L_x$), as highlighted by a dashed green box in the $u$-field.
Analysis on such a dataset not only avoids uncertainties due to Taylor's hypothesis approximation \citep{dennis2008,del2009,wu2012}, but also permits identification of these superstructures directly from an instantaneous flow field of a high $Re_{\tau}$ boundary layer (where superstructures are statistically significant).
The latter represents another unique feature of the present study, and overcomes the limitations experienced by past experimental studies \citep{liu2001,guala2006,balakumar2007}, which were restricted to isolating superstructure characteristics based on Fourier-filtering, or Proper orthogonal decomposition (POD)-based decomposition of ensemble/time-averaged statistics.
The accuracy of these LFOV PIV datasets have been firmly established in appendix 1 ($\S$\ref{app1}), which compares the premultiplied 1-D spectra obtained from the present data, with those acquired via multiwire anemometry published previously \citep{winter2015,baidya2017}.
Readers can also refer to the same appendix section for details associated with the computation of the velocity spectra from PIV flow fields, which is relevant to the analysis presented ahead in the paper.

The fourth and final PIV dataset comprises of relatively smaller flow fields in the $x$--$z$ plane (in terms of $\delta$-scaling), and is hence referred to as simply the PIV dataset.
This was acquired at $U_{\infty}$ $\approx$ 20\;m{s$^{-1}$}, close to the downstream end of the test section ($x$ $\approx$ 21\;m from the trip), where $\delta$ $\approx$ 0.3\;m, yielding a high $Re_{\tau}$ $\approx$ 14500.
The full velocity field captured in this experiment was also made possible by using the same eight PCO 4000 cameras, arranged in two vertical rows of four cameras each, to capture the significantly thicker boundary layer (refer to figures 1-2 of \citealp{charitha2014}).
This limits the streamwise extent of the flow field to $x$ $\sim$ 2$\delta$ in this case, and is hence not used for identifying the turbulent superstructures in instantaneous fields, but rather used to compute the two-point correlations of $u$- and $w$-fluctuations along the $z$-direction (limited to the inner-region).
It is owing to this reason that only a part of the full flow field ($x$ ${\times}$ $z$ $\sim$ 2${\delta}$ $\times$ 0.4$\delta$), from this dataset, has been considered in the present study.
The image pairs from all four PIV datasets were processed via an in-house PIV package developed by the Melbourne group \citep{charitha2014}, with the final window sizes (${\Delta}{x^+}$,${\Delta}{z^+}$) used for processing given in table \ref{tab1}.
Interested readers may refer to the cited references for further details about the experimental setup and methodology adopted for acquiring these datasets.

The fifth dataset, which is at the highest $Re_{\tau}$ $\sim$ $\mathcal{O}$(10$^6$), was acquired at the Surface Layer Turbulence and Environmental Science Test (SLTEST) facility in the salt flats of western Utah.
The data is acquired from a spanwise and wall-normal array of 18 sonic anemometers (Campbell Scientific CSAT3) arranged in an `L'-shaped configuration (refer to figure 1 of \citealp{hutchins2012}).
While the full dataset comprises of continuous measurements of all three velocity components as well as the temperature at the SLTEST site over a duration of nine days, here we limit our attention solely to one hour of data associated with near-neutral (i.e. near canonical) atmospheric boundary layer conditions \citep{hutchins2012}.
These conditions were confirmed based on estimation of the Monin–Obukhov similarity parameter, determined on averaging across the 10 sonic anemometers placed along the spanwise array, at a fixed distance from the wall ($z$ $\approx$ 2.14\;m).
For the present analysis, we are solely interested in the $u$- and $w$-fluctuations measured synchronously by the nine sonic anemometers on the wall-normal array, which were placed between 1.42\;m $\le$ $z$ $\le$ 25.69\;m with logarithmic spacing. 
Mean streamwise velocity measurements reported by \citet{hutchins2012} confirm that all these $z$-locations fall within the log-region of the atmospheric boundary layer.
This data is also used here to compute the two-point correlations of $u$- and $w$-fluctuations along the $z$-direction, for comparison with those obtained from the PIV datasets acquired in the laboratory.

\subsection{Methodology employed to identify and extract turbulent superstructures}
\label{extract}

In the present study, we are interested in computing conditional statistics of the velocity fluctuations associated with the superstructures, identified from the individual flow fields in the LFOV PIV dataset.
To identify these structures, we need to first define what we mean by a superstructure, for which we draw inspiration from past studies that have investigated these motions based on 3-D instantaneous flow fields \citep{hutchins2007,dennis2011Part2,lee2011}. Those studies, as noted by \cite{smits2011}, refer to superstructures as ``\emph{very long, meandering, features consisting of narrow regions of low-streamwise-momentum fluid flanked by regions of higher-momentum fluid}'', that ``\emph{have also been observed in the logarithmic and wake regions of wall flows.}''
Here, for the purpose of analyzing 2-D velocity fields, we define superstructures as very large-scale motions that persist spatially with coherent regions of streamwise velocity, and account for a significant fraction of the streamwise turbulent kinetic energy.
Identifying these structures from the PIV field, hence, requires establishing logical thresholds to the geometric and kinematic properties of the fluctuating $u$-field \citep{hwang2018,charitha2020}.
For this, we consider previous findings and adopt the following thresholds:
\begin{enumerate}
\item |$u$($x$,$z$)| $>$ $\sqrt{\overline{{u^2}(z)}}$ \citep{liu2001}, where $\sqrt{\overline{{u^2}(z)}}$ is the root-mean-square of the $u$-fluctuations at $z$.
\item streamwise extent, $L_x$ $>$ 3$\delta$ \citep{guala2006,balakumar2007}.
\item wall-normal extent should at least span across 2.6$\sqrt{Re_{\tau}}$ $\lesssim$ $z^+$ $\lesssim$ 0.5$Re_{\tau}$ \citep{guala2006,hutchins2007,balakumar2007,deshpande2021model}.
\end{enumerate}
In the process of identifying a superstructure, the threshold associated with the streamwise turbulent kinetic energy (i.e. (i)) is considered first before applying thresholds associated with the geometric extent ((ii) and (iii)).
With regards to criteria (ii), we acknowledge that past studies investigating 3-D instantaneous flow fields \citep{hutchins2007,lee2011,dennis2011Part2} have found superstructures to be as long as 10-20$\delta$.
However, statistical analysis based on 1-D one-/two-point correlations \citep{guala2006,hutchins2007,balakumar2007,deshpande2021model} suggests these structures have relatively modest lengths (on average), between 3-6$\delta$.
Considering that the present analysis is also limited to 2-D flow fields, we adapt the estimates from past statistical analyses and consider $u$-structures with streamwise extent, $L_x$ $>$ 3$\delta$ as superstructures.
Figure \ref{fig3}(d) gives an example of a -$u$ superstructure identified and extracted by the algorithm ($u|_{SS}$), based on the aforementioned thresholds from the full flow field depicted in figure \ref{fig3}(b) (highlighted by the dashed green box).
Streamwise extent/length of the identified structures ($L_{x}$) is judged based on the length of a rectangular bounding box (along $x$) that fully encompasses the identified structure.
Our superstructure identification algorithm extracts the rectangular 2-D flow field within this box to conduct further conditional analysis associated with the superstructures.
Although the choice of a rectangular box inevitably also brings in some part of the flow not associated with a superstructure, it only forms a minor part ($\sim$20\%) of the bounding box, suggesting conditional statistics can be predominantly associated with the superstructures.
Interested readers are referred to the supplementary document provided along with this manuscript, which provides a step-by-step description of the superstructure identification and extraction procedure from a 2-D PIV flow field.
\begin{figure}
   \captionsetup{width=1.0\linewidth}
  \centerline{\includegraphics[width=1.0\textwidth]{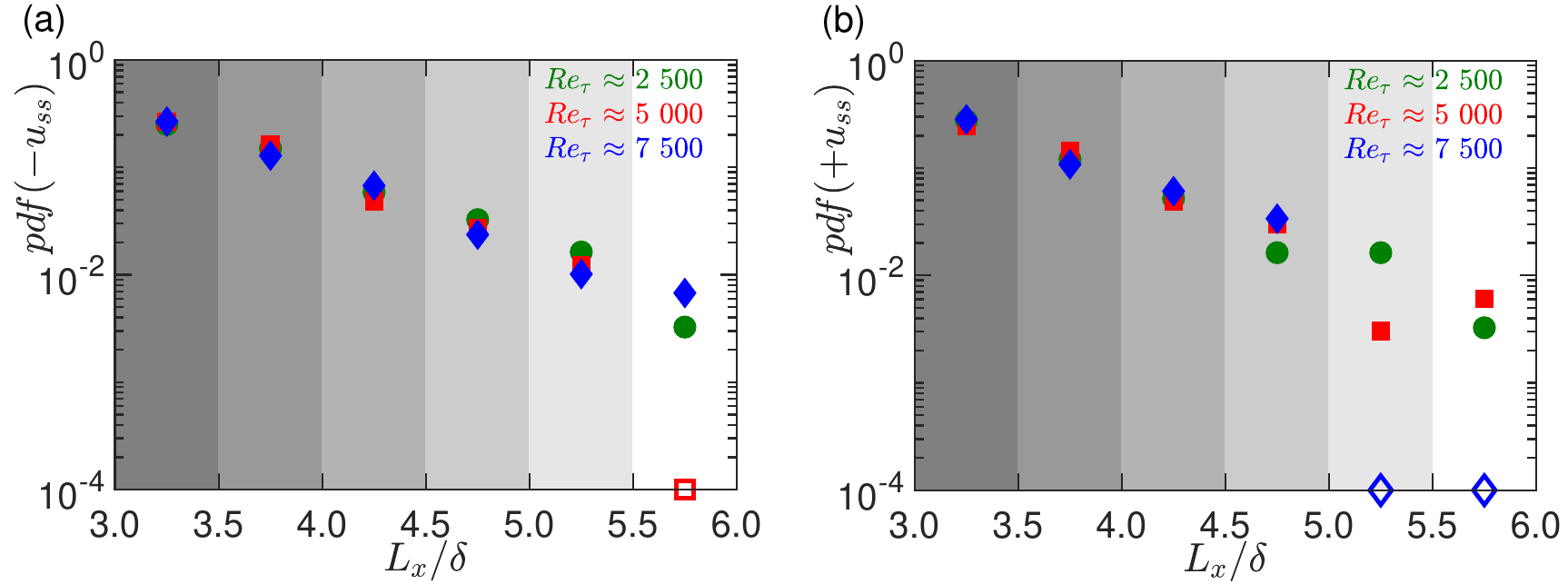}}
  \caption{Probability density function (\emph{pdf}) of the lengths of the large and intense, (a) low and (b) high streamwise momentum motions detected by the superstructure extraction algorithm in PIV flow fields of various $Re_{\tau}$.
	Background shading indicates the bin sizes used to estimate the \emph{pdf}, for which the total number of detected superstructures (i.e. addition of $+u_{ss}$ and $-u_{ss}$) was used for normalization.
Empty symbols indicate zero probability for the respective bin.}
\label{fig4}
\end{figure}

Besides identifying a superstructure, which is indicated by a dashed green box in figures \ref{fig3}(b,c), the algorithm also identifies a region of same length$\times$height as the green box but not associated with a superstructure ($u|_{noSS}$).
The $u|_{noSS}$ flow field region is allocated by the algorithm in the same wall-normal range as $u|_{SS}$, but in a different streamwise location within the PIV image that does not satisfy criteria (i-ii) defined above, thereby ensuring it doesn't overlap with $u|_{SS}$.
This practice of extracting $u|_{noSS}$, from the same PIV fields used to extract $u|_{SS}$ and of the same size as that of $u|_{SS}$, is conducted across all three LFOV datasets to form a set of $u|_{noSS}$ and $u|_{SS}$ of equal ensembles.
Conditional statistics are computed and compared from both $u|_{SS}$ and $u|_{noSS}$, with the latter considered to confirm that the trends depicted by the former are not an artefact of aliasing or insufficient ensembling/noise.

The superstructure extraction algorithm described above, identified superstructures of both +$u$ and -$u$ signatures, of varying lengths, from the three LFOV PIV datasets. 
A summary of their streamwise extents is presented in the form of a probability distribution function ($pdf$) plot in figure \ref{fig4}.
The plot is obtained by sorting the identified $u$-motions into bins of width 0.5$\delta$ (between 3.0:0.5:6.0), based on their respective lengths ($L_{x}$). 
The population associated with each bin is then normalized by the total number of $-u$ and $+u$ superstructures identified by the algorithm (for $L_{x}$ $>$ 3$\delta$), which is then plotted in the figure.
It can be noted from the plots that the $pdf$s do not change significantly with $Re_{\tau}$ for structures of lengths, $L_{x}$ $<$ 5$\delta$. 
It is only when $L_{x}$ is increased significantly ($>$ 5$\delta$) that notable differences appear for different $Re_{\tau}$.
For example, no $-u$ or $+u$-structures are identified in certain PIV datasets while in others, the probability is low.
Further, the logarithmic scaling of the vertical axis of the plots reveals that the population density decreases near exponentially as the criteria (ii) to identify a superstructure (i.e. minimum length, $L_x$) is increased.
The effect of increasing the minimum streamwise extent of a $u$-structure to qualify as a superstructure, on the conditionally averaged statistics, has been documented in figure \ref{fig15} in Appendix 2 ($\S$\ref{app2}).
Given that an increase in $L_x$ does not change the scaling behaviour, but significantly reduces the convergence of the conditioned statistics (due to fewer ensembles), reinforces the choice of $L_x$ $\gtrsim$ 3$\delta$ in criteria (ii) discussed above.

\section{Mean statistics}
\label{mean_results}

\begin{figure}
   \captionsetup{width=1.0\linewidth}
  \centerline{\includegraphics[width=1.0\textwidth]{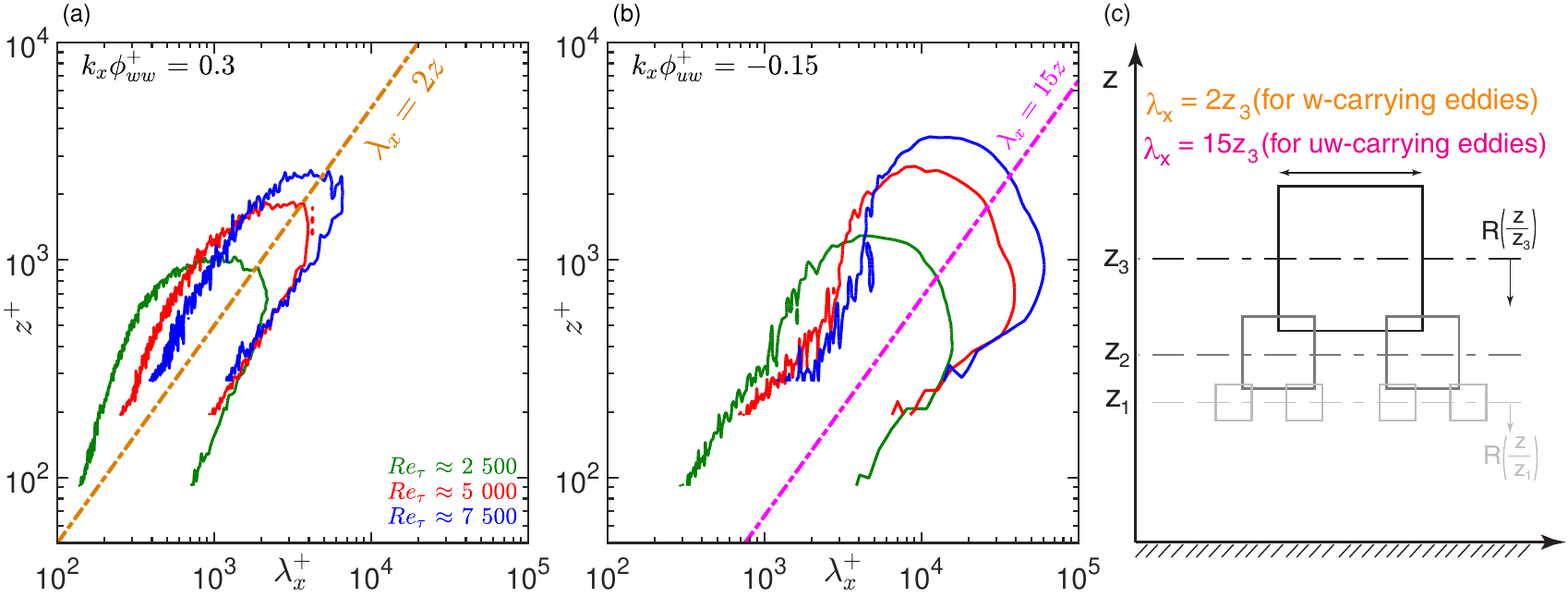}}
  \caption{Iso-contours of the premultiplied streamwise 1-D (a) energy spectra of $w$-fluctuations and (b) co-spectra of the Reynolds shear stress plotted against $z^+$ and ${\lambda}^{+}_{x}$, computed from the LFOV PIV dataset at various $Re_{\tau}$.  
	{\color{Golden}Dash-dotted golden} and {\color{magenta}magenta} lines represent the relationships ${\lambda}_{x}$ $\approx$ 2$z$ and ${\lambda}_{x}$ $\approx$ 15$z$, respectively following \citet{baidya2017}.
(c) Schematic of representative $w$ and $uw$-carrying eddies centred at various distances from the wall ($z_{r}$) in the log region, with light to dark shading used to suggest an increase in $z_{r}$.  
$R_{ww}$($z/{z_{r}}$) and $R_{uw}$($z/{z_{r}}$) respectively represent the vertical coherence of the $w$- and $uw$-carrying eddies centred at $z_{r}$.}	
\label{fig5}
\end{figure}

\begin{figure}
   \captionsetup{width=1.0\linewidth}
  \centerline{\includegraphics[width=1.0\textwidth]{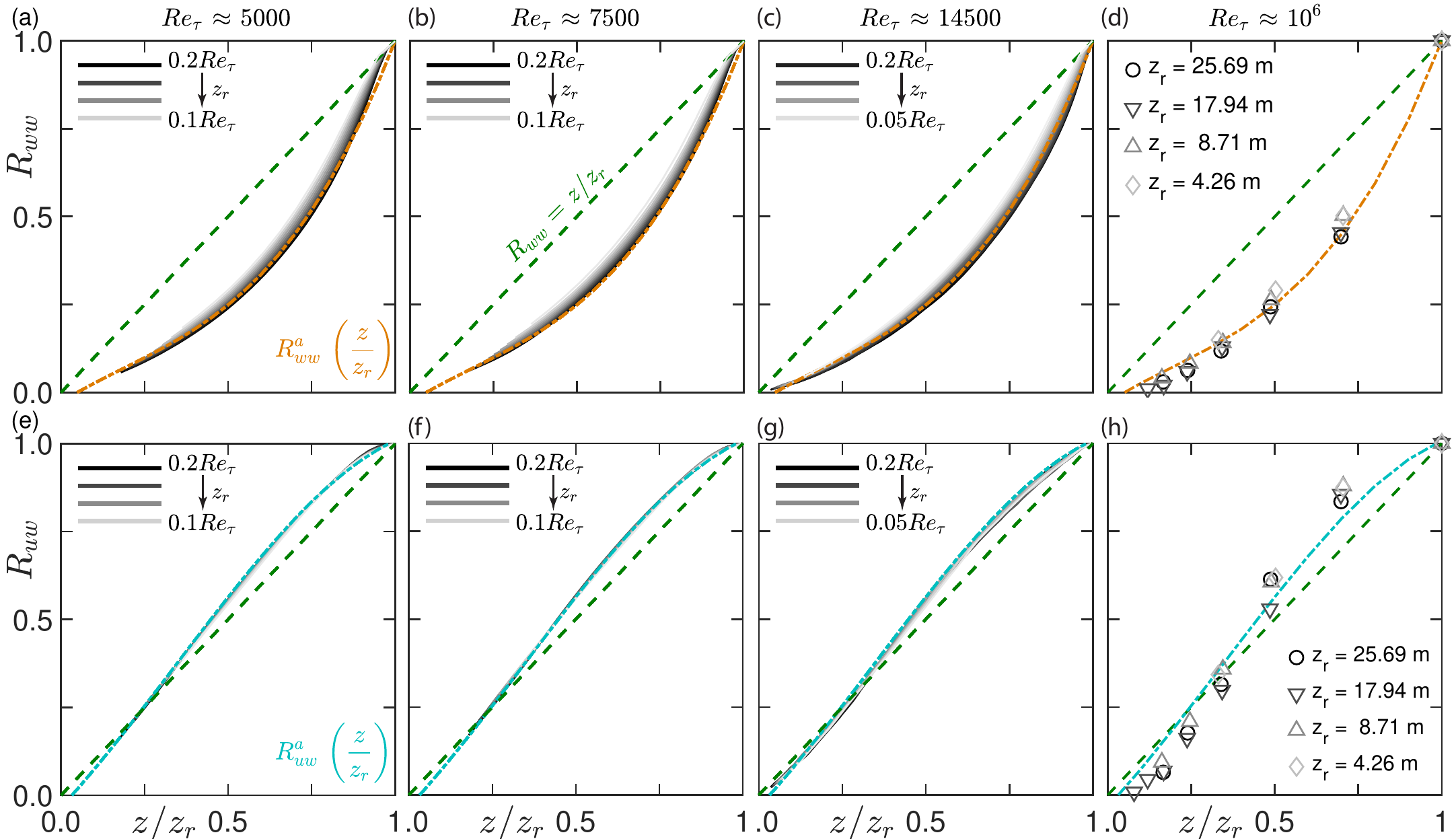}}
  \caption{(a-d) Cross-correlation of $w$-fluctuations measured at $z$ and $z_{r}$, normalized by $\overline{w^2}$($z_{r}$) for various $z_{r}$ in the log-region. 
	(e-h) Cross-correlation between $u$($z_{r}$) and $w$($z$), normalized by $\overline{uw}$($z_{r}$) for the same $z_{r}$ as in (a-d). .
(a,b,e,f) are estimated from the LFOV PIV datasets while (c,g) have been computed from the PIV dataset. 
In case of the SLTEST dataset in (d,h), $z_{r}$ listed in the legend corresponds to the 9$^{th}$, 8$^{th}$, 6$^{th}$ and 4$^{th}$ sonic positioned from the ground.
{\color{OliveGreen}Dashed green} line corresponds to the linear relationship, $z/{z_{r}}$ while {\color{Golden}dash-dotted golden} and {\color{Teal}teal} lines correspond to the least-squares fit ${R^{a}_{ww}}$ and ${R^{a}_{uw}}$ defined in (\ref{eq2}), respectively.}	
\label{fig6}
\end{figure}

Before investigating the conditionally averaged statistics associated with the superstructures, it is worth revisiting the scaling behaviour of the mean statistics, against which the former would be compared.
Here, the mean statistics have been obtained by averaging across all 3000 flow fields, and considering the entire 12$\delta$ long flow fields in case of the LFOV PIV datasets.
In the present study, since we are primarily interested in the $w$-velocity behaviour associated with superstructures, we investigate the mean spatial coherence of the $w$-carrying eddies in the log-region of a high $Re_{\tau}$ boundary layer.
We look at the spatial coherence in both the streamwise (figure \ref{fig5}) as well as wall-normal direction (figure \ref{fig6}), for both the $w$-fluctuations and the Reynolds shear stress ($uw$).
Previous investigations on the vertical coherence have been rare compared to the streamwise coherence, particularly for the log-region of a high $Re_{\tau}$ boundary layer, owing to the lack of large-scale PIV experiments of the kind utilized here.
This makes the present investigation (figure \ref{fig6}) unique by itself.

Figures \ref{fig5}(a,b) depict the iso-contours of the premultiplied spectrogram of the $w$-velocity and the Reynolds shear stress respectively, computed from the three LFOV PIV datasets.
These are plotted as a function of ${\lambda}^{+}_{x}$ and $z^+$.
The iso-contours for the $w$-velocity spectrograms can be seen centred around the linear ($z$-)scaling indicated by ${\lambda}_{x}$ $=$ 2$z$ for all $Re_{\tau}$, which is consistent with previous observations in the literature \citep{baidya2017}.
Similarly, the iso-contours for the Reynolds shear stress spectrograms also follow a linear scaling (${\lambda}_{x}$ $=$ 15$z$) for all $Re_{\tau}$, again consistent with the literature \citep{baidya2017}.
This analysis not only validates the spectra estimated from the LFOV PIV, but also assists with the construction of a simplified 2-D conceptual picture of the $w$- and $uw$-carrying eddies in the log-region of a high $Re_{\tau}$ boundary layer (figure \ref{fig5}c).
Here, based on the $z$-scaling exhibited by the data, the lengths (${\lambda}_{x}$) of the $w$- and $uw$-carrying eddies have been defined as 2$z_{r}$ and 15$z_{r}$ respectively, where $z_{r}$ represents the distance of the eddy centre from the wall.
This scaling confirms the association of these $w$- and $uw$-carrying eddies with Townsend's attached eddy hierarchy, according to which attached eddies scale with $z_{r}$ \citep{townsend1976,baidya2017,deshpande2021}.

While both these linear scalings, which represent the streamwise coherence of the $w$- and $uw$-carrying eddies, are well accepted in the literature, not much is known about the vertical/wall-normal coherence of the same eddying motions at high $Re_{\tau}$.
Several previous studies \citep{bellot1963,tritton1967,sabot1973,hunt1987,liu2001,sillero2014} have investigated their vertical coherence in low $Re_{\tau}$ canonical wall flows via traditional two-point correlations, providing interesting insights on their scaling.
Here, we are inspired by one such interesting result reported in the seminal work of \citet{hunt1988}, based on high $Re_{\tau}$ unstably stratified atmospheric boundary layer data, who found the two-point correlation coefficients given by:
\begin{equation}
\label{eq1}
\begin{aligned}
{R_{ww}}{\bigg(}\frac{z}{z_r}{\bigg)} = \frac{\overline{{w(z)}{w(z_r)}}}{\overline{{w^{2}}(z_{r})}} \hspace{1cm}  \textrm{and} \hspace{1cm} {R_{uw}}{\bigg(}\frac{z}{z_r}{\bigg)} = \frac{\overline{{w(z)}{u(z_r)}}}{\overline{{uw}(z_{r})}},
\end{aligned}
\end{equation}
to be a function of $({z/z_r})$.
Here, $z_{r}$ acts as the reference wall-normal location fixed in the log-region, thereby making $R_{ww}$ (or $R_{uw}$) representative of the vertical coherence of the eddy centred at $z_{r}$.
It should be noted here that these correlation functions are different from the conventionally used two-point correlations (which consider normalization by the root-mean-square of velocity at both $z$ and $z_{r}$), and hence their values aren't restricted between -1 and 1.
 Equations (\ref{eq1}), however, are ideally suited for the present study, which tests the self-similarity (i.e. $z$-scaling) of the vertical coherence of the momentum carrying eddies.
We compute these correlations for the four high $Re_{\tau}$ boundary layer datasets considered and plot them in figure \ref{fig6}, for various $z_{r}$ restricted to the log-region.
It can be clearly observed that $R_{uw}$ curves for varying $z_{r}$ and $Re_{\tau}$ collapse over one another (represented by a line in teal colour based on least-squares fit), suggesting $Re_{\tau}$-invariance via $z$-scaling of the vertical coherence of $uw$-carrying motions.
On the other hand, the collapse in the $R_{ww}$ curves is not as good for the relatively low $Re_{\tau}$ cases ($<$ 7500), but certainly gets better for the very high $Re_{\tau}$ atmospheric data (figure \ref{fig6}d).
This case has a significantly thicker log-region than the boundary layers generated in the lab, suggesting the influence of the wall behind the relatively poor collapse of $R_{ww}$ at low $Re_{\tau}$.
Accordingly, the $z$-scaling of the $R_{ww}$ curves has been represented by the golden lines in figures \ref{fig6}(a-d) (obtained by a least-squares fit), which are consistent with $R_{ww}$ curves in figure \ref{fig6}(d), as well as $R_{ww}$ estimated farthest from the wall ($z^{+}_{r}$ $\approx$ 0.2$Re_{\tau}$) in figures \ref{fig6}(a-c).
The analytical expressions associated with these golden and teal lines are:
\begin{equation}
\label{eq2}
\begin{aligned}
{R^{a}_{ww}}{\bigg(}\frac{z}{z_r}{\bigg)} &= 1.007{{{\bigg(}\frac{z}{z_r}{\bigg)}}^3} - 0.56{{{\bigg(}\frac{z}{z_r}{\bigg)}}^2} + 0.58{{{\bigg(}\frac{z}{z_r}{\bigg)}}} - 0.027, \hspace{0.2cm}  \textrm{and}\\
{R^{a}_{uw}}{\bigg(}\frac{z}{z_r}{\bigg)} &= - 0.65{{{\bigg(}\frac{z}{z_r}{\bigg)}}^3} + 0.65{{{\bigg(}\frac{z}{z_r}{\bigg)}}^2} + 1.03{{{\bigg(}\frac{z}{z_r}{\bigg)}}} - 0.03.
\end{aligned}
\end{equation}
The fact that $R_{ww}$ and $R_{uw}$ are solely a function of $z/{z_{r}}$ represents geometric self-similarity in the vertical coherence of the $w$- and $uw$-carrying inertial eddies, reaffirming their association with Townsend's attached eddies.
The analytical forms in (\ref{eq2}) can thus be used in data-driven coherent structure-based models \citep{deshpande2021model} to simulate high $Re_{\tau}$ boundary layers (such as atmospheric surface layers).
It is worth noting that the collapse in the $R_{uw}$ and $R_{ww}$ curves, observed in figure \ref{fig6}, does not exist for $w$- and $uw$-carrying eddies centred far outside the log-region of the boundary layer (i.e. $z_r$ $>$ 0.2$\delta$; not shown here), which may be due to the growing influence of the turbulent/non-turbulent interface in the outer-region \citep{charitha2014}.
Investigations for $z_{r}$ below the log-region, however, were not possible owing to insufficient data points captured by the LFOV PIV.

\section{Conditionally averaged statistics associated with superstructures}
\label{CondAvg_results}

\begin{figure}
   \captionsetup{width=1.0\linewidth}
  \centerline{\includegraphics[width=1.0\textwidth]{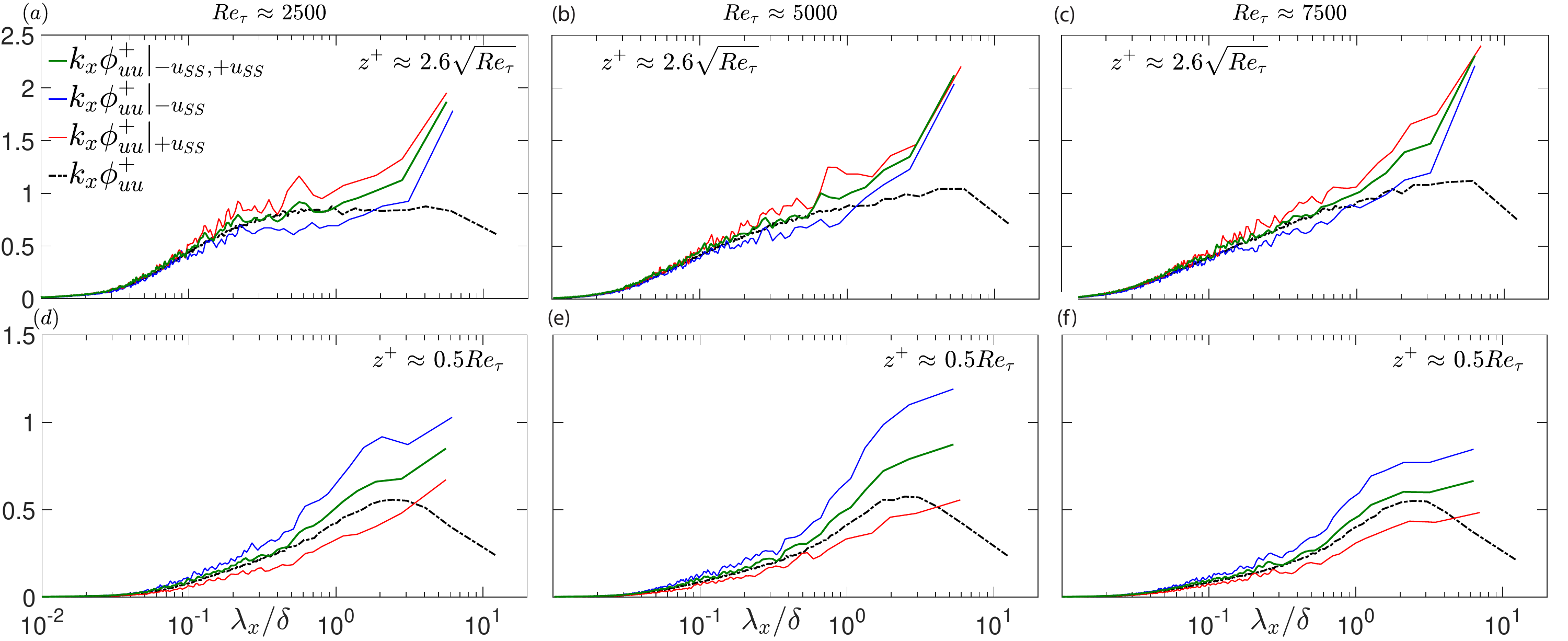}}
  \caption{(a-f) Premultipled 1-D spectra of the $u$-fluctuations plotted versus ${{\lambda}_{x}}/{\delta}$ at (a-c) $z^+$ $\approx$ 2.6$\sqrt{Re_{\tau}}$ and (d-f) $z^+$ $\approx$ 0.5$Re_{\tau}$ for LFOV PIV data at $Re_{\tau}$ $\approx$ (a,d) 2500, (b,e) 5000 and (c,f) 7500. 
Dashed black lines correspond to the mean spectra obtained by ensembling across 3000 PIV images of the full flow field. 
While, the solid {\color{blue}blue} and {\color{red}red} lines represent conditional spectra computed from the $u$-flow field extracted based on identification of a -${u_{ss}}$ and +${u_{ss}}$, respectively. 
The spectra in {\color{OliveGreen}green} is computed by ensembling across both -${u_{ss}}$ and +${u_{ss}}$.}	
\label{fig7}
\end{figure}

With the scaling behaviour of the mean statistics established in $\S$\ref{mean_results}, we progress next towards analyzing the conditionally averaged statistics (spectra and correlations) associated with superstructures.
Figure \ref{fig7} plots the conditionally averaged, premultiplied $u$-spectra computed from the extracted flow fields associated with superstructures (figure \ref{fig3}d), from the three LFOV PIV datasets.
The spectra are plotted for $z^+$ $\approx$ 2.6$\sqrt{Re_{\tau}}$ and 0.5$Re_{\tau}$, and estimated individually from the extracted flow fields associated with low-momentum (${{k_{x}}{{\phi}^{+}_{uu}}}|_{-{u_{ss}}}$; {\color{blue}in blue}) and high-momentum superstructures (${{k_{x}}{{\phi}^{+}_{uu}}}|_{+{u_{ss}}}$; {\color{red}in red}).
Also plotted is the conditionally averaged spectra considering both -$u_{ss}$ and +$u_{ss}$ (${{k_{x}}{{\phi}^{+}_{uu}}}|_{{-{u_{ss}}},{+{u_{ss}}}}$; {\color{OliveGreen}in green}), which is compared against the mean $u$-spectra shown in figures \ref{fig14}(a-c).
A noteworthy observation from the conditionally averaged spectra (${{k_{x}}{{\phi}^{+}_{uu}}}|_{{-{u_{ss}}},{+{u_{ss}}}}$) is the enhanced large-scale energy (${\lambda}^{+}_{x}$ $\gtrsim$ 10$^4$) seen for all three $Re_{\tau}$ cases.
These enhanced energy levels are due to the significant streamwise turbulent kinetic energy associated with the superstructures, which is captured in the extracted flow fields and averaged across fewer ensembles, than those used for obtaining the mean spectra.
\begin{figure}
   \captionsetup{width=1.0\linewidth}
  \centerline{\includegraphics[width=1.0\textwidth]{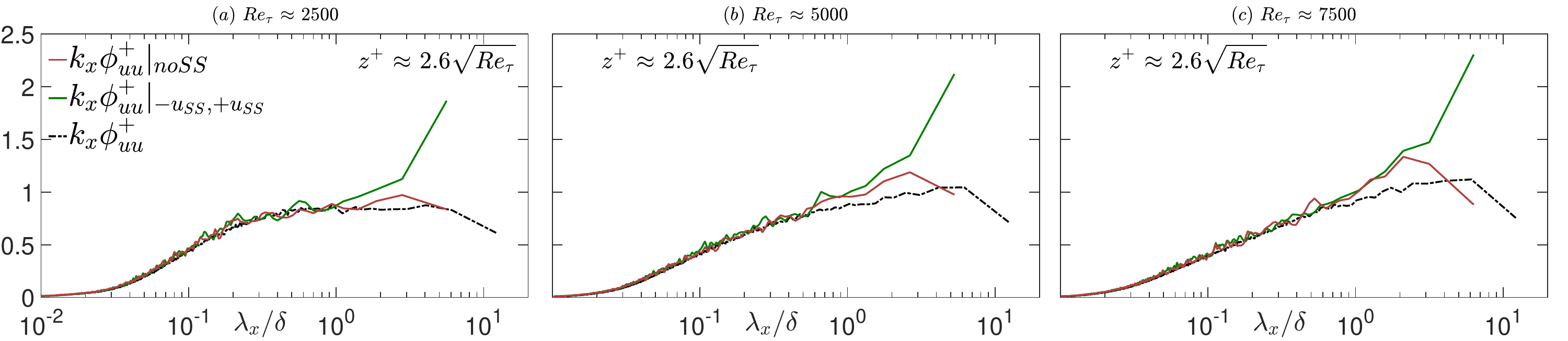}}
  \caption{Premultiplied 1-D spectra of the $u$-fluctuations plotted versus ${{\lambda}_{x}}/{\delta}$ at $z^+$ $\approx$ 2.6$\sqrt{Re_{\tau}}$ for $Re_{\tau}$ $\approx$ (a) 2500, (b) 5000 and (c) 7500.
	The mean spectra estimated from the full flow field (in black lines) is ensembled across all 3000 fields.
	While, the conditional spectra corresponds to extracted flow fields ($\sim$ 300) of the same length$\times$height associated (in {\color{OliveGreen}green}) and not associated (in {\color{brown}brown}) with the superstructures.}	
\label{fig8}
\end{figure}
To confirm that these trends are not an artefact of aliasing or ensembling, figure \ref{fig8} compares the conditionally averaged spectra associated with superstructures ({\color{OliveGreen}green boxes} in figure \ref{fig3}c) with that not associated with the superstructures ({\color{brown}brown boxes} in figure \ref{fig3}c).
Given that both the conditional spectra are estimated from the same number of extracted flow fields, of the same length$\times$height, the enhanced energy in the largest scales for ${{k_{x}}{{\phi}^{+}_{uu}}}|_{{-{u_{ss}}},{+{u_{ss}}}}$ (compared to ${{k_{x}}{{\phi}^{+}_{uu}}}|_{noSS}$) can be unambiguously associated with the turbulent superstructures.
These trends give us confidence regarding the efficacy of the superstructure extraction algorithm.
Also, they indicate that the scalings observed from the conditionally averaged $u$-, $w$-statistics can be associated with the constituent motions of superstructures.
This is one of the advantages of analyzing very-large-scale motions based on extraction of instantaneous flow fields (present study), as compared with the much simpler approach of Fourier filtering (past studies).

Another interesting observation from the conditional spectra for low- and high-momentum motions, ${{k_{x}}{{\phi}^{+}_{uu}}}|_{-{u_{ss}}}$ and ${{k_{x}}{{\phi}^{+}_{uu}}}|_{+{u_{ss}}}$, is their starkly different behaviour in the lower portion of the log-region (figures \ref{fig7}a-c) and outside of it (figures \ref{fig7}d-e).
While ${{k_{x}}{{\phi}^{+}_{uu}}}|_{+{u_{ss}}}$ $>$ ${{k_{x}}{{\phi}^{+}_{uu}}}|_{-{u_{ss}}}$ for $z^+$ $\approx$ 2.6$\sqrt{Re_{\tau}}$, it is vice versa for $z^+$ $\approx$ 0.5$Re_{\tau}$, which is in accordance with previous observations made by \citet{hutchins2007}.
The study reported that the turbulent structures associated with +$u_{ss}$ are more energetic than those associated with -$u_{ss}$, in the lower part of the log-region.
This behaviour, however, is reversed in the upper part of the log-region and beyond, which is consistent with our observations from figure \ref{fig7}, reaffirming confidence in the extracted flow fields.

\begin{figure}
   \captionsetup{width=1.0\linewidth}
  \centerline{\includegraphics[width=1.0\textwidth]{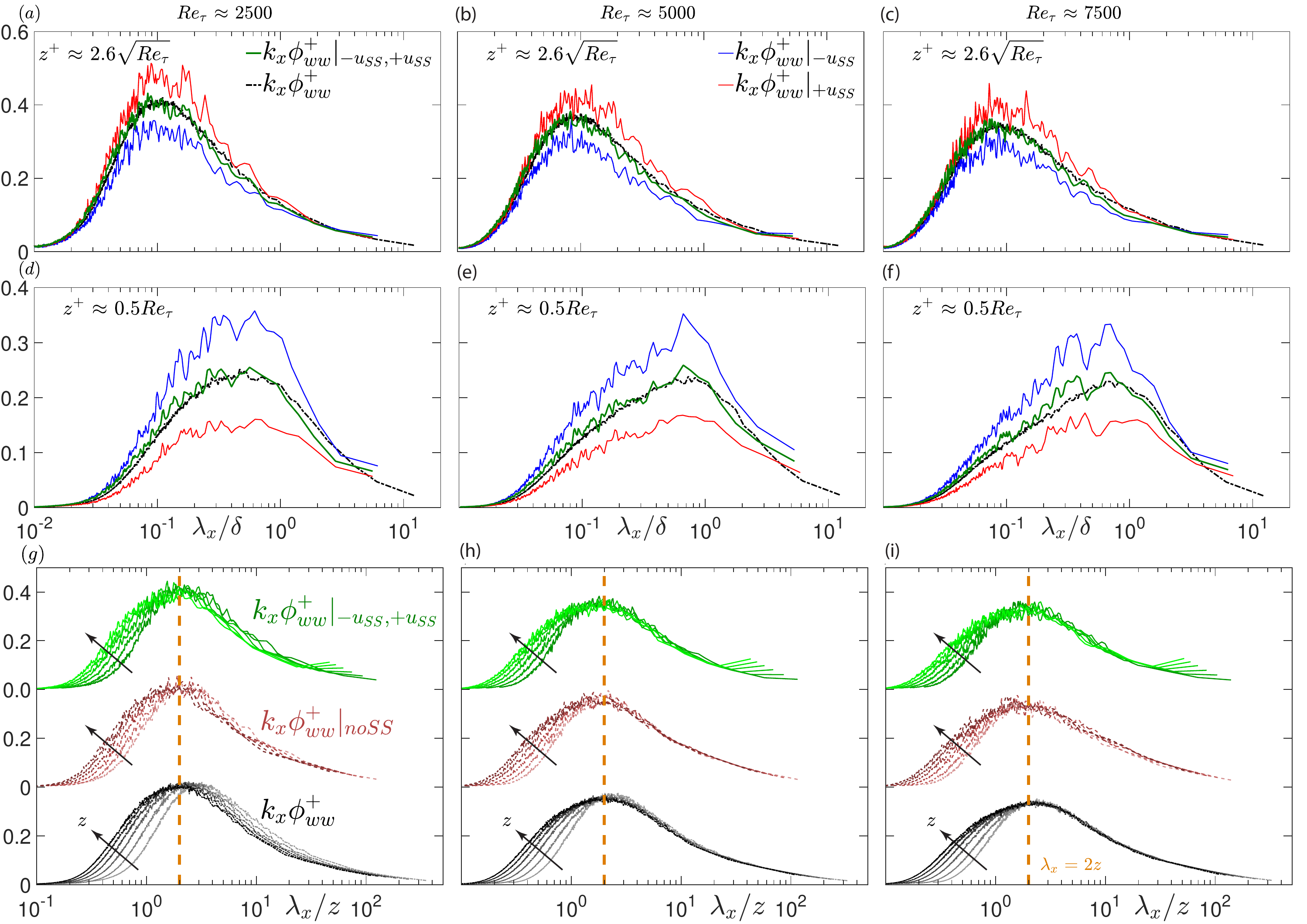}}
  \caption{(a-f)Premultipled 1-D spectra of the $w$-fluctuations plotted versus ${{\lambda}_{x}}/{\delta}$ at (a-c) $z^+$ $\approx$ 2.6$\sqrt{Re_{\tau}}$ and (d-f) $z^+$ $\approx$ 0.5$Re_{\tau}$ for large FOV PIV data at $Re_{\tau}$ $\approx$ (a,d,g) 2500, (b,e,h) 5000 and (c,f,i) 7500. 
Dashed black lines correspond to the mean spectra obtained by ensembling across 3000 PIV images of the full flow field. 
While, the solid {\color{blue}blue} and {\color{red}red} lines represent conditional spectra computed from the extracted $w$-flow fields associated with -${u_{ss}}$ and +${u_{ss}}$, respectively. 
The spectra in {\color{OliveGreen}green} corresponds to all extracted $w$-flow fields, associated with both -${u_{ss}}$ and +${u_{ss}}$.
(g-i)	Premultiplied 1-D spectra of $w$-velocity plotted vs ${{\lambda}_{x}}/{z}$ at various $z^+$ within the log region (2.6$\sqrt{Re_{\tau}}$ $\lesssim$ $z^+$ $\lesssim$ 0.15$Re_{\tau}$). 
Colour coding is the same as that defined for plots (a-f). 
Line plots in {\color{brown}brown} correspond to conditional spectra ${{k_{x}}{{\phi}^{+}_{ww}}}|_{noSS}$ computed from extracted flow fields of the same length$\times$height as the ${{k_{x}}{{\phi}^{+}_{ww}}}|_{{-{u_{ss}}},{+{u_{ss}}}}$, but not associated with the superstructures.
{\color{Golden}Dashed golden} lines represent the linear scaling, ${\lambda}_{x}$ $=$ 2$z$.}
\label{fig9}
\end{figure}

With the efficacy of the superstructure extraction algorithm now established, we shift our focus to the statistical quantity of primary interest: conditionally averaged, premultplied $w$-spectra associated with superstructures.
Figures \ref{fig9}(a-f) plot ${{k_{x}}{{\phi}^{+}_{ww}}}|_{-{u_{ss}}}$, ${{k_{x}}{{\phi}^{+}_{ww}}}|_{+{u_{ss}}}$ and ${{k_{x}}{{\phi}^{+}_{ww}}}|_{{-{u_{ss}}},{+{u_{ss}}}}$ for $z^+$ $\approx$ 2.6$\sqrt{Re_{\tau}}$ and 0.5$Re_{\tau}$ computed from all three LFOV PIV datasets.
Here again, ${{k_{x}}{{\phi}^{+}_{ww}}}|_{+{u_{ss}}}$ $>$ ${{k_{x}}{{\phi}^{+}_{ww}}}|_{-{u_{ss}}}$ for $z^+$ $\approx$ 2.6$\sqrt{Re_{\tau}}$, and vice versa for $z^+$ $\approx$ 0.5$Re_{\tau}$, which is consistent with the behaviour noted for the $u$-spectra in figure \ref{fig7}.
Interestingly ${{k_{x}}{{\phi}^{+}_{ww}}}|_{{-{u_{ss}}},{+{u_{ss}}}}$, which represents $w$-energy associated with both -$u_{ss}$ and +$u_{ss}$, can be seen overlapping with the mean $w$-spectra, for both $z^+$ and all three $Re_{\tau}$ considered in figure \ref{fig9} (except at the largest ${\lambda}^{+}_{x}$).
This suggests that the $w$-carrying eddies within the superstructures conform to the inertia-dominated $z$-scaled eddies predominant in the log-region (figure \ref{fig5}).
To test the $z$-scaling characteristics of ${{k_{x}}{{\phi}^{+}_{ww}}}|_{{-{u_{ss}}},{+{u_{ss}}}}$, we plot it for various $z^+$ corresponding to the log-region in figures \ref{fig9}(g-i).
Remarkably, ${{k_{x}}{{\phi}^{+}_{ww}}}|_{{-{u_{ss}}},{+{u_{ss}}}}$ exhibits $z$-scaling behaviour similar to the mean spectra for ${\lambda}_{x}$ $\gtrsim$ 2$z$, across all three $Re_{\tau}$, suggesting that the geometrically self-similar attached eddies are likely the constituent motions of the superstructures.
Further, the peak of ${{k_{x}}{{\phi}^{+}_{ww}}}|_{{-{u_{ss}}},{+{u_{ss}}}}$ also scales with ${\lambda}_{x}$ $=$ 2$z$, which is consistent with the mean spectra.

\begin{figure}
   \captionsetup{width=1.0\linewidth}
  \centerline{\includegraphics[width=1.0\textwidth]{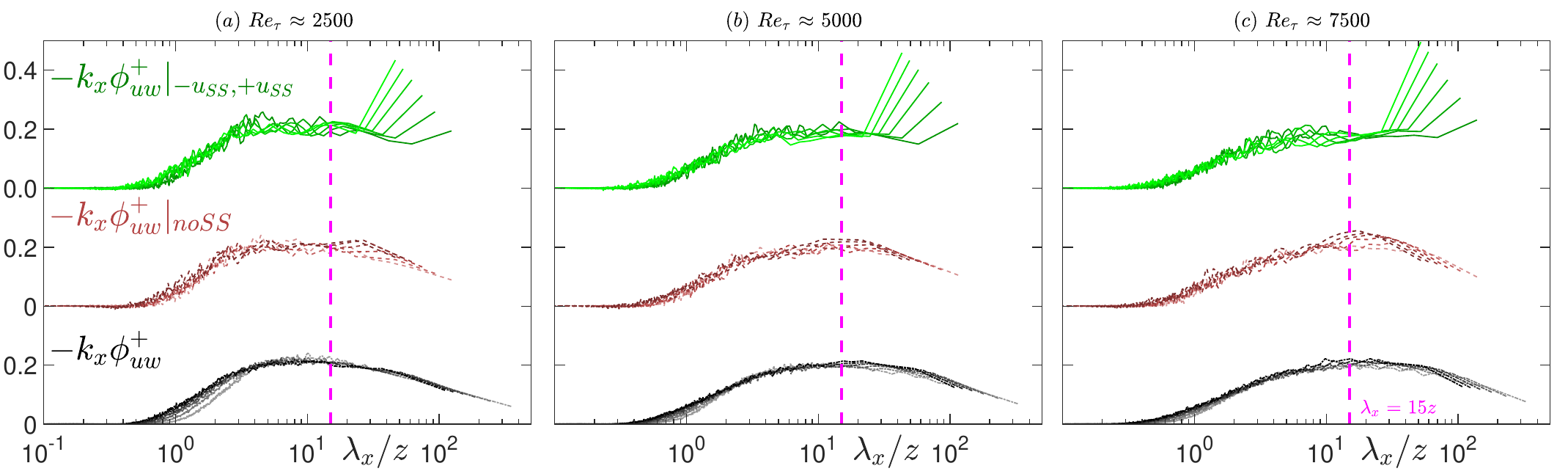}}
  \caption{(a-c) Premultiplied 1-D co-spectra of the Reynolds shear stress plotted vs ${{\lambda}_{x}}/z$ at various $z^+$ within the log region (2.6$\sqrt{Re_{\tau}}$ $\lesssim$ $z^+$ $\lesssim$ 0.15$Re_{\tau}$). 
Dashed black lines correspond to the mean spectra obtained by ensembling across 3000 PIV images of the full flow field. 
The spectra in {\color{OliveGreen}green} corresponds to all extracted $u$,$w$-flow fields, associated with both -${u_{ss}}$ and +${u_{ss}}$. 
Line plots in {\color{brown}brown} correspond to conditional spectra ${{k_{x}}{{\phi}^{+}_{uw}}}|_{noSS}$ computed from extracted flow fields of the same length$\times$height as the ${{k_{x}}{{\phi}^{+}_{uw}}}|_{{-{u_{ss}}},{+{u_{ss}}}}$, but not associated with the superstructures.
{\color{magenta}Dashed magenta} lines represent the linear scaling, ${\lambda}_{x}$ $=$ 15$z$.}	
\label{fig10}
\end{figure}

A unique observation associated with ${{k_{x}}{{\phi}^{+}_{ww}}}|_{{-{u_{ss}}},{+{u_{ss}}}}$, as compared to other spectra plotted in figure \ref{fig9}, is the slightly enhanced energy at the very-large ${\lambda}_{x}$.  
There are two possible interpretations to this very-large-scale peak, with the first (and the most likely one) associated with the preservation of the covariance tensor (i.e. $\overline{uw}$) for the large scales. 
Based on the definition of the structure parameter, which is essentially the normalized form of $\overline{uw}$ that tends to $\sim$ $\mathcal{O}$(1) for the large scales, conditioning the $w$-field based on a large-scale $u$-structure is bound to yield non-zero $w$-energy in the large-scales, as a mathematical artefact.
The other possible interpretation is the presence of a physically long $w$-eddy in the flow field, on conditioning of the flow based on a turbulent superstructure.
Considering the instantaneous $w$-field discussed previously in figures \ref{fig2}(c,e), however, the second scenario seems very less likely, and this is reaffirmed via further analysis in $\S$\ref{physical}.
The fact that the very-large-scale peak is not an artefact of aliasing or experimental noise is confirmed by considering ${{k_{x}}{{\phi}^{+}_{ww}}}|_{noSS}$ plotted for the same $z^+$-range in figures \ref{fig9}(g-i).
Since ${{k_{x}}{{\phi}^{+}_{ww}}}|_{noSS}$ is estimated based on the same number of ensembles and length$\times$height of the extracted flow fields as ${{k_{x}}{{\phi}^{+}_{ww}}}|_{{-{u_{ss}}},{+{u_{ss}}}}$, the former would have also had enhanced energy levels at large ${\lambda}^{+}_{x}$, if these peaks were owing to aliasing or noise.

Another way to reaffirm the close association between the geometrically self-similar attached eddies and the superstructures is by analyzing the Reynolds shear stress.
Previous studies \citep{liu2001,guala2006,balakumar2007} have found superstructures to carry a significant proportion of the Reynolds shear stress in the log-region, and this is confirmed by the comparison of the Reynolds shear stress co-spectra presented in figure \ref{fig10} at various $Re_{\tau}$.
This figure plots the conditionally averaged co-spectra associated with the superstructures (${{k_{x}}{{\phi}^{+}_{uw}}}|_{{-{u_{ss}}},{+{u_{ss}}}}$), the co-spectra not associated with the superstructures (${{k_{x}}{{\phi}^{+}_{uw}}}|_{noSS}$), as well as the ensemble-averaged co-spectra (${{k_{x}}{{\phi}^{+}_{uw}}}$) at various $z^+$ within the log-region.
Similar to that noted for the $w$-spectra, the $z$-scaling observed in the ensemble-averaged co-spectra (${{\lambda}_{x}}$ = 15$z$; figure \ref{fig5}b) is also noted for the ${{k_{x}}{{\phi}^{+}_{uw}}}|_{{-{u_{ss}}},{+{u_{ss}}}}$, confirming our claim that the self-similar motions coexist within the superstructure region.
This comparison between ${{k_{x}}{{\phi}^{+}_{uw}}}|_{{-{u_{ss}}},{+{u_{ss}}}}$ and ${{k_{x}}{{\phi}^{+}_{uw}}}$ also showcases the significance of analyzing the very-large-scale motions by extracting instantaneous flow fields, than using pure Fourier filtering.
While the latter is simpler to execute, it doesn't present the `full physical picture' associated with the very-large-scale motions.
It is only after extraction of the instantaneous flow fields at high $Re_{\tau}$ that the present study can confirm the $z$-scaling characteristics associated with the constituent motions of the superstructures (${{k_{x}}{{\phi}^{+}_{uw}}}|_{{-{u_{ss}}},{+{u_{ss}}}}$).
In figure \ref{fig10}, again, high energy levels can be noted in ${{k_{x}}{{\phi}^{+}_{uw}}}|_{{-{u_{ss}}},{+{u_{ss}}}}$ at very large ${{\lambda}_{x}}$, the magnitude of which is much greater than the energy levels for ${{k_{x}}{{\phi}^{+}_{uw}}}|_{noSS}$ and ${{k_{x}}{{\phi}^{+}_{uw}}}$ at the same ${{\lambda}_{x}}$.
Further analysis is presented in $\S$\ref{physical} to reaffirm that these peaks do not represent very-large-scale $w$-motions existing in the physical flow field.

\begin{figure}
   \captionsetup{width=1.0\linewidth}
  \centerline{\includegraphics[width=1.0\textwidth]{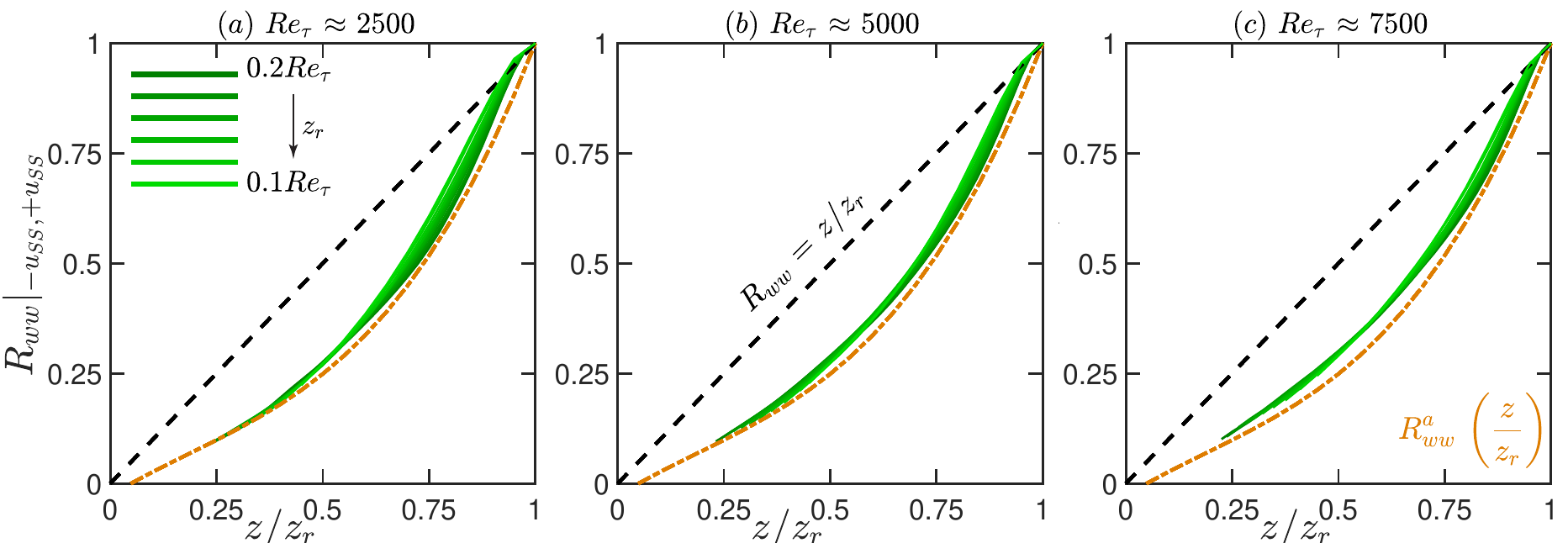}}
  \caption{(a-f) Conditionally averaged correlations between $w$-fluctuations at $z$ and $z_{r}$, normalized by ${\overline{w^2}}({z_r})$ for various $z_{r}$. 
	The correlations have been computed from the extracted $w$-flow fields associated with both -${u_{ss}}$ and +${u_{ss}}$. 
	Dashed black line corresponds to the linear relationship, $z/{z_{r}}$ while {\color{Golden}dashed dotted golden} line corresponds to $R^{a}_{ww}$ defined in (\ref{eq2}).}	
\label{fig11}
\end{figure}

While the conditional 1-D spectra brings out the geometric characteristics of the constituent motions along the streamwise direction, the former can be understood for the wall-normal direction by computing the two-point correlations ($R_{ww}$; (\ref{eq1})) for the extracted flow fields.
Figure \ref{fig11} plots ${R_{ww}}|_{{-{u_{ss}}},{+{u_{ss}}}}$, i.e. the two-point correlations computed from the $w$-fluctuations associated with both -$u_{ss}$ and +$u_{ss}$, for $z_{r}$ limited to the log-region.
These are estimated for all three LFOV PIV datasets and compared with the least-squares fit (given by (\ref{eq2})) estimated from the mean statistics (plotted with a golden line).
Consistent with our observations based on the mean statistics in figure \ref{fig6}, the collapse in the ${R_{ww}}|_{{-{u_{ss}}},{+{u_{ss}}}}$ curves is not very good at low $Re_{\tau}$ but improves significantly at $Re_{\tau}$ $\approx$ 7500.
Interestingly, however, ${R_{ww}}|_{{-{u_{ss}}},{+{u_{ss}}}}$ curves estimated at all $Re_{\tau}$ are close to the empirically obtained least-squares fit.
Hence, investigation of the vertical coherence of the $w$-carrying eddies (associated with superstructures) also indicates that the geometrically self-similar eddies coexist within superstructure region, in line with interpretations based on figures \ref{fig9} and \ref{fig10}. 
While the present study lacks the analysis to investigate the spanwise coherence of the constituent motions, consideration of the present findings in light of the recent knowledge on the log region \citep{hwang2018,deshpande2020,deshpande2021,deshpande2021model} suggests that they would likely exhibit self-similar characteristics along the span as well.
Notably, \citet{deshpande2020,deshpande2021model} found that the spanwise extent of the wall-coherent, intermediate-scaled motions (${\lambda}_{x}$ $\lesssim$ 4$\delta$) varies self-similarly with respect to their streamwise extent, which directly corresponds to the scale-range associated with the constituent motions of the superstructures.

\subsection{Physical interpretations and discussions on the conditionally averaged statistics}
\label{physical}

Here, we discuss the physical interpretation of the conditionally-averaged spectra presented in figures \ref{fig7}-\ref{fig10}, and how it advances our understanding of the constituent motions forming the turbulent superstructures. 
Given the geometry of individual $w$-eddies does not physically conform with the very-large-scale peaks noted in ${{k_{x}}{{\phi}^{+}_{ww}}}|_{{-{u_{ss}}},{+{u_{ss}}}}$ and ${{k_{x}}{{\phi}^{+}_{uw}}}|_{{-{u_{ss}}},{+{u_{ss}}}}$ (discussed previously based on figures \ref{fig2}c,e), these peaks are likely an artefact of the preservation of the covariance tensor, which is a property of the Fourier transform.
However, the non-zero correlation between $u$ and $w$-fluctuations, at large ${\lambda}_{x}$, has often been misinterpreted to be representative of instantaneous $w$-features physically as long as the superstructures (as also highlighted by \citealp{lozano2012} and \citealp{sillero2014}), especially when one analyzes it from the perspective of the structure parameter (${\sim}{\mathcal{O}}$(1) for large $\lambda_{x}$).
Here, we prove from our analysis that this interpretation is incorrect.
If one observes $w|_{ss}$ plotted in figure \ref{fig3}(e), which is conditioned with respect to a $-u$ superstructure, it is clear there are no long and energetic $w$-features extending beyond 3$\delta$.
To the best of the authors' knowledge, energetic $w$-features of such long streamwise extents have never been noted in instantaneous flow fields, and their absence can also be confirmed from the negligible energy in the 1-D $w$-spectra plotted in figure \ref{fig5}(a), or in the literature \citep{baidya2017}.
Absence of very-long ($\gtrsim$3$\delta$) $w$-features also means there are no very-large-scale Reynolds shear stress-carrying motions in the instantaneous flow \citep{lozano2012,sillero2014}.
Such misinterpretations are the source for the long-standing contradictions between the attached eddy hypothesis and past studies \citep{guala2006,balakumar2007,wu2012} investigating the Reynolds shear stress co-spectra (refer $\S$\ref{intro}), which we attempt to clarify here.

\begin{figure}
   \captionsetup{width=1.0\linewidth}
  \centerline{\includegraphics[width=1.0\textwidth]{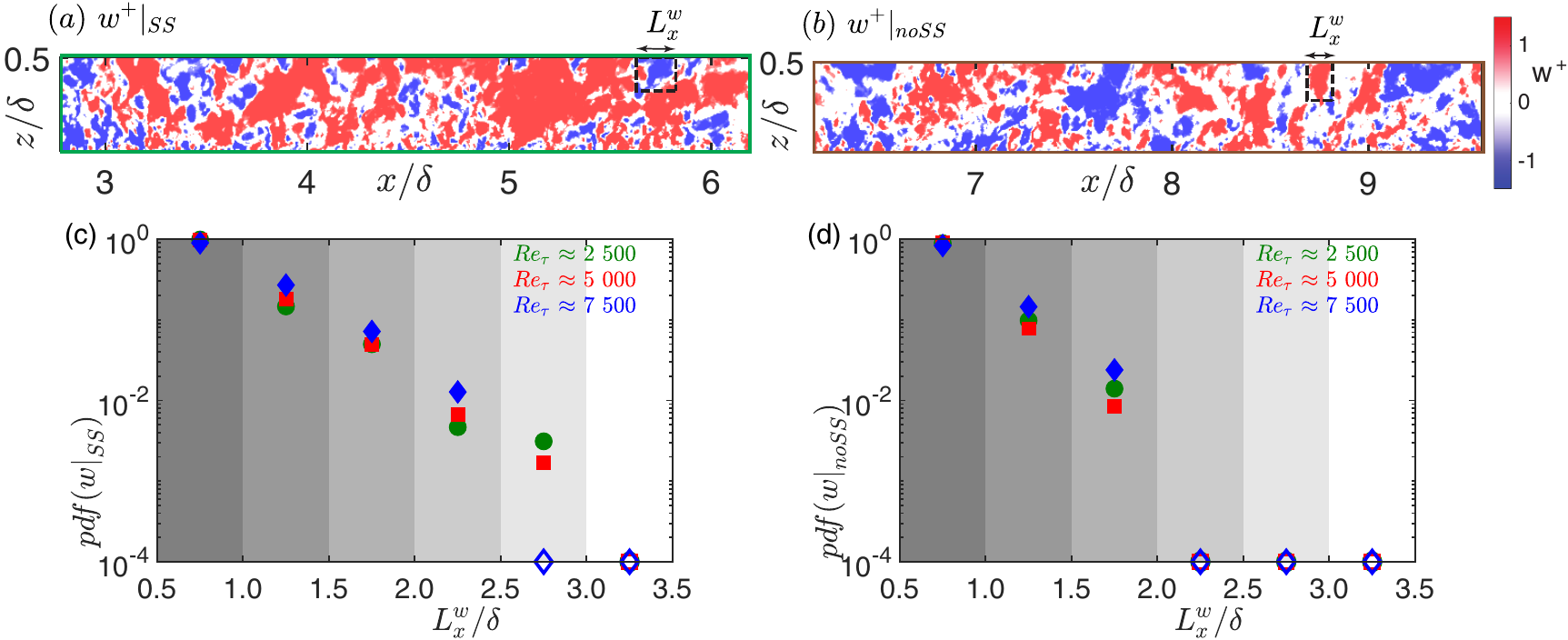}}
  \caption{(a,b) Examples of intense $w$-fluctuations ($|{w_{SS}}|$, $|{w_{noSS}}|$ $>$ 1.3$\sqrt{{\overline{w^2}}(z)}$) present within flow fields associated (a) with superstructures ($w|_{SS}$) and (b) not associated with superstructures ($w|_{noSS}$). 
	The $w|_{SS}$ and $w|_{noSS}$ flow fields used as examples in (a,b) essentially correspond to the extracted fields shown in figures \ref{fig3}(e,g).
(c,d) Probability density function (\emph{pdf}) of the lengths ($L^{w}_{x}$) of intense (c) $w|_{SS}$ and (d) $w|_{noSS}$ motions extracted from the corresponding flow fields at various $Re_{\tau}$.
Background shading indicates the bin size used to estimate the $pdf$, for which the total number of detected $w|_{SS}$ and $w|_{noSS}$ were used for normalization.
Empty symbols indicate zero probability for the respective bin.}	
\label{fig12}
\end{figure}

To reaffirm that the very-large-scale peaks in ${{k_{x}}{{\phi}^{+}_{ww}}}|_{{-{u_{ss}}},{+{u_{ss}}}}$ do not correspond with very long and energetic $w$-features in the instantaneous flow, figure \ref{fig12} analyzes the streamwise extent of $w$-eddies ($L^{w}_{x}$) in the extracted $w|_{SS}$ and $w|_{noSS}$ fields.
For this analysis, the same algorithm is deployed to identify and characterize the $w$-eddies, as used to identify and extract superstructures in the $u$-field (refer $\S$\ref{extract} and the supplementary document).
Figures \ref{fig12}(a,b) represent the same $w|_{SS}$ and $w|_{noSS}$ fields as in figures \ref{fig3}(e,g), but only consider motions with strong fluctuations (i.e. $|{w_{SS}}|$, $|{w_{noSS}}|$ $>$ 1.3$\sqrt{{\overline{w^2}}(z)}$).
This threshold is based on \citet{dennis2011Part2} and assists with identification and extraction of individual, energetic $w$-eddies.
Figures \ref{fig12}(c,d) present the probability distribution functions of the streamwise extents of the $w$-eddies identified within $w|_{SS}$ and $w|_{noSS}$ flow fields, extracted across all three PIV datasets.
The $pdfs$ confirm that the streamwise extent of $w$-eddies is limited to $\lesssim$ 3$\delta$ across both $w|_{SS}$ and $w|_{noSS}$.
This wavelength range closely corresponds with the geometrically self-similar hierarchy of eddies exhibiting distance-from-the-wall scaling in figure \ref{fig9}, reaffirming the key finding of this study, based on direct analysis of the physical flow field.
Although not shown here, a similar analysis on the Reynolds shear stress-carrying eddies also yields the same conclusion, reinforcing our earlier statements on the interpretation of ${{k_{x}}{{\phi}^{+}_{uw}}}|_{{-{u_{ss}}},{+{u_{ss}}}}$.
The analysis also confirms that energetic $w$-eddies do not physically extend along $x$, as long as the superstructures ($>3{\delta}$), meaning the only possible way of observing a physically long $w$-feature is when the individual $w$-eddies align along the $x$-direction.
Indeed, the $w|_{SS}$ flow field indicates a much more closely-packed/clustered organization of the individual $w$-eddies, compared to $w|_{noSS}$, in the $x$-direction (figure \ref{fig12}a,b).
However, since the present analysis uses snapshot 2-D PIV data, this study cannot definitively comment on the dynamics associated behind the formation of superstructures.
But, the conditional analysis presented in this section does lend empirical support in favour of the formation of superstructures, via streamwise concatenation of the intermediate-scaled eddies \citep{adrian2000}.
Interested readers are referred to the supplementary document, where we have utilized a simplified coherent structure-based model (i.e. the attached eddy model), to demonstrate a statistically plausible scenario of self-similar eddies aligning in $x$ to `form' a superstructure.

The present results are also consistent with the conclusions of \citet{lozano2012}, who found large-scale Reynolds shear stress-carrying structures to be essentially a concatenation of smaller $uw$-carrying eddies, having lengths $\sim$3 times their height.
This also clarifies the contradiction in the literature on the `active'/`inactive' status of the very-large-scale $u$-motions (i.e. superstructures).
Given there are no very-large-scale $w$- (and consequently $uw$-) features in the instantaneous flow, the superstructures are indeed inactive as per the definition of Townsend \citep{deshpande2021}.
Present evidence indicates that the superstructures comprise of several $z$-scaled $w$-carrying (i.e. active) motions, which explains the past empirical observations of superstructures carrying significant Reynolds shear stress.

\section{Concluding remarks}

The present study analyzes large-scale PIV datasets, acquired in moderate to high $Re_{\tau}$ turbulent boundary layers (2500 $\lesssim$ $Re_{\tau}$ $\lesssim$ 7500), to investigate the constituent motions of the turbulent superstructures.
Considering that superstructures are statistically significant only at $Re_{\tau}$ $\gtrsim$ 2000 \citep{hutchins2007}, the present datasets (providing sufficient scale separation) are ideally suited to identify superstructures and analyze their constituent motions.
These unique datasets accurately capture the inertia-dominated instantaneous $u$- and $w$-fluctuations across a large streamwise wall-normal plane, extending up to 12$\delta$ in the $x$-direction. 
This facilitates a comprehensive investigation of the horizontal (via 1-D spectra) as well as vertical coherence (via two-point correlations) of the Reynolds shear stress-carrying eddies coexisting in the log-region, which are responsible for the momentum transfer in a high $Re_{\tau}$ boundary layer \citep{baidya2017,deshpande2021}.
The statistics bring out the geometric self-similarity of these energetically significant eddies, which complements the well-established knowledge on the self-similarity exhibited by the wall-parallel velocity components in a canonical flow \citep{baars2017,hwang2018,deshpande2020}.
We note that this motivates undertaking similar investigations of the momentum and heat flux in thermally stratified wall-bounded flows at high $Re_{\tau}$ (for example atmospheric boundary layers), which can likely assist with coherent structure-based modelling of these practically relevant flows.

\begin{figure}
   \captionsetup{width=1.05\linewidth}
  \centerline{\includegraphics[width=1.0\textwidth]{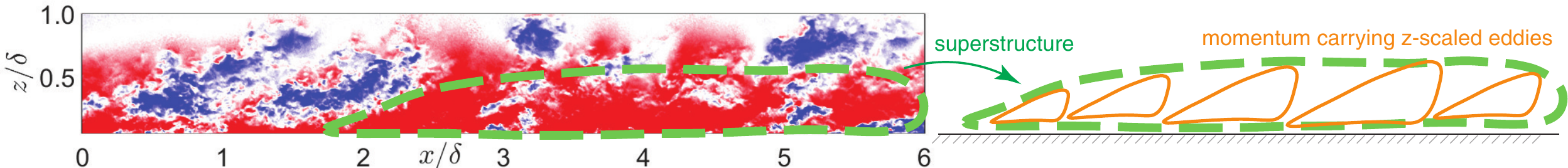}}
  \caption{Conceptual representation of the main conclusion of this study: $z$-scaled eddies are likely the constituent motions forming the turbulent superstructures.}
\label{fig13}
\end{figure}

The empirically derived scaling behaviour observed from these mean statistics (spectra and correlations) provide a benchmark for comparing and contrasting with the conditionally averaged statistics, associated with the turbulent superstructures.
Such conditional statistics are made possible by the large-scale PIV flow fields, which permit identification of the superstructures directly from instantaneous flow fields.
These statistics present a comprehensive picture of the superstructures, in comparison to the limited information available based on modal decompositions, used often in past studies (such as Fourier filtering, etc.).
Considering the ambiguity involved while interpreting the smaller constituent motions from a $u$-flow field, the present study adopts the approach of investigating the $w$-fluctuations within the superstructure region, to understand its constituent motions.
Notably, the conditional streamwise $w$- and $uw$-spectra exhibit the classical $z$-scaling (${\lambda}_{x}$ = 2$z$; ${\lambda}_{x}$ = 15$z$) in the intermediate scale range \citep{baidya2017}, clearly suggesting that geometrically self-similar eddies co-exist within the superstructure region (represented schematically in figure \ref{fig13}).
The same conclusion is demonstrated through the conditional two-point $w$-correlations, along the vertical direction, which also exhibit self-similar scaling similar to that noted for the mean flow.
Investigations of these kinds are only possible on analyzing instantaneous flow fields, highlighting the uniqueness of the present large-scale high $Re_{\tau}$ PIV dataset.

The argument regarding the self-similar motions, as the likely constituent motions of the turbulent superstructures, is reaffirmed by analyzing the geometry and population of individual $w$-eddies associated with these very-large-scaled structures.
The maximum streamwise extent of the energetic $w$-eddies was found limited to $\lesssim$ 3$\delta$ within the superstructures, similar to that noted outside a superstructure, and conforming to the self-similar hierarchy of scales.
The same analysis also revealed the spatial organization of these constituent $w$-eddies within the superstructures, which is consistent with the streamwise concatenation argument of forming superstructures.
This also helps clarify longstanding contradiction in the literature on the active/inactive behaviour of the superstructures \citep{guala2006,balakumar2007,wu2012}.
Since there are no very-large-scale $w$- and $uw$-features in the instantaneous flow, the superstructures are indeed inactive per the definition of Townsend \citep{deshpande2021}.
However, the study finds that superstructures comprise of several $z$-scaled Reynolds shear stress-carrying (i.e. active) motions \citep{lozano2012}, which explains the past empirical observations of these very-large-scaled motions carrying significant Reynolds shear stress.

The present study concludes that superstructures are an assemblage of the attached eddy hierarchy in the streamwise wall-normal plane, hinting at a well-defined spatial organization of the attached eddies.
This contradicts the original hypothesis of \citet{townsend1976}, per which attached eddies are randomly distributed in the flow domain, suggesting the need to revisit the hypothesis (this has also been tested based on synthetic flow fields and presented in the supplementary document).
The present empirical findings, specifically the $Re_{\tau}$-invariance of the vertical coherence of inertial eddies ($R_{ww}$, $R_{uw}$), can also be used to further improve coherent structure-based models, such as the attached eddy model \citep{marusic2019}.
This is possible through extending the data-driven approach proposed recently in \citet{deshpande2021model}, by defining the geometry of the representative eddies based on the least-squares fits presented in (\ref{eq2}).
The present findings would also benefit the attached eddy model by acting as empirical evidence, to model superstructures as clusters of self-similar (attached) eddies, organized along the streamwise direction.

\section*{Acknowledgements}

The authors wish to acknowledge the Australian Research Council for financial support and are grateful to Prof. N. Hutchins for insightful discussions regarding this work. R.D. acknowledges partial financial support by the University of Melbourne through the Melbourne Postdoctoral Fellowship.
The authors also thank an anonymous referee for highlighting the significance of the structure parameter while performing conditional analysis for the large-scales.

\section*{Declaration of Interests} 

The authors report no conflict of interest.

\section*{Appendix 1: Comparisons of energy spectra obtained from LFOV PIV with published hotwire data}
\label{app1}

\begin{figure}
   \captionsetup{width=1.0\linewidth}
  \centerline{\includegraphics[width=1.0\textwidth]{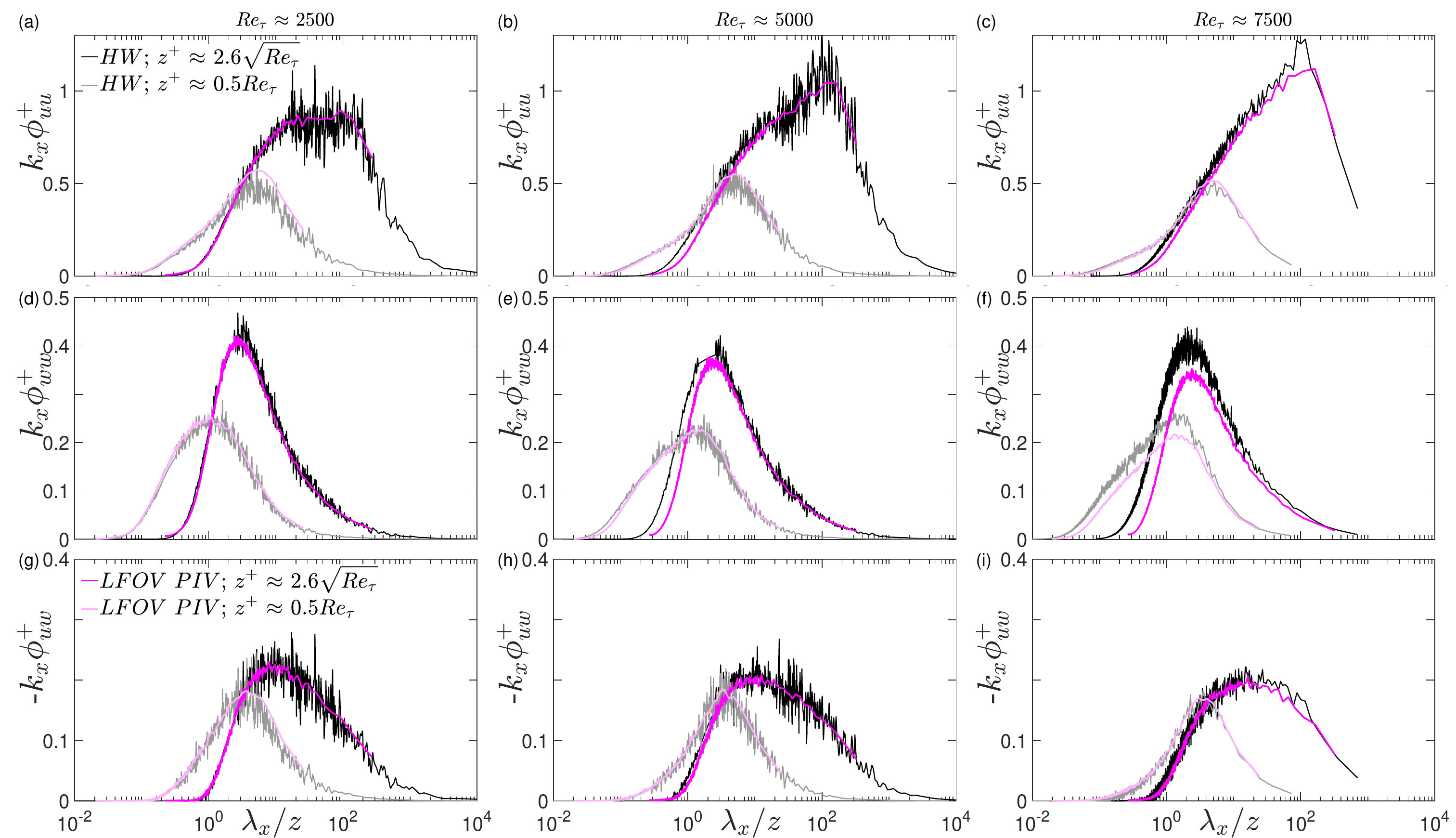}}
  \caption{Premultiplied 1-D spectra of (a-c) $u$-fluctuations, (d-f) $w$-fluctuations and (g-i) co-spectra of the Reynolds shear stress at $Re_{\tau}$ $\approx$ (a,d,g) 2500, (b,e,h) 5000 and (c,f,i) 7500. 
Lines in {\color{magenta}magenta} correspond to the spectra estimated from the LFOV PIV datasets documented in table \ref{tab1}, while the lines in black correspond to the same computed from the multiwire data acquired by \citet{baidya2017} (at $Re_{\tau}$ $\approx$ 2500 and 5000) and \citet{winter2015} (at $Re_{\tau}$ $\approx$ 7500), in the same wind tunnel facility.
	Dark shaded lines correspond to spectra at $z^{+}$ $\approx$ 2.6$\sqrt{Re_{\tau}}$, while light shaded lines correspond to $z^+$ $\approx$ 0.5${Re_{\tau}}$.}	
\label{fig14}
\end{figure}

Figure \ref{fig14} plots the premultiplied streamwise 1-D spectra of the streamwise and wall-normal velocity components, as well as the Reynolds shear stress obtained from the LFOV PIV dataset.
Throughout this study, we are limited to studying the 1-D velocity spectrum, which is a function of the streamwise wavelengths, ${\lambda}_{x}$ and distance from the wall, $z$. 
The spectrum at $z$ is computed by first extracting the velocity fluctuations measured at various streamwise locations at the given height, from a PIV flow field.
The extracted 1-D array is zero-padded with sufficient elements to avoid uncertainties due to a `tight' box size.
Then, a Hamming window function \citep{bendat2011} is considered to reduce spectral leakage, which is non-zero for the length/array size of the original PIV flow field (along $x$) and zero outside (to correspond with the zero padding).
Next, the array associated with the Hamming window function is multiplied on an element-by-element basis with the zero-padded 1-D array, comprising the velocity fluctuations.
Fourier transform is then computed on this 1-D array to get the streamwise 1-D spectrum of the velocity component.
This procedure was adopted while computing each spectra (ensemble-averaged as well as the conditionally-averaged spectra) presented in this study.
While the Hamming window was used predominantly in this study, this spectra was also compared with spectra computed using Dirichlet or Welch window functions \citep{bendat2011} for certain cases. 
No difference/effect was noted in the spectra on changing the window functions.
Also, since the experimental dataset is acquired by a snapshot PIV technique (i.e. not time-resolved), there was no overlap in the instantaneous flow fields acquired consecutively by the PIV system.
Hence, Welch's method of overlapping signals, to improve statistical convergence, couldn't be employed here.

\begin{figure}
   \captionsetup{width=1.0\linewidth}
  \centerline{\includegraphics[width=1.0\textwidth]{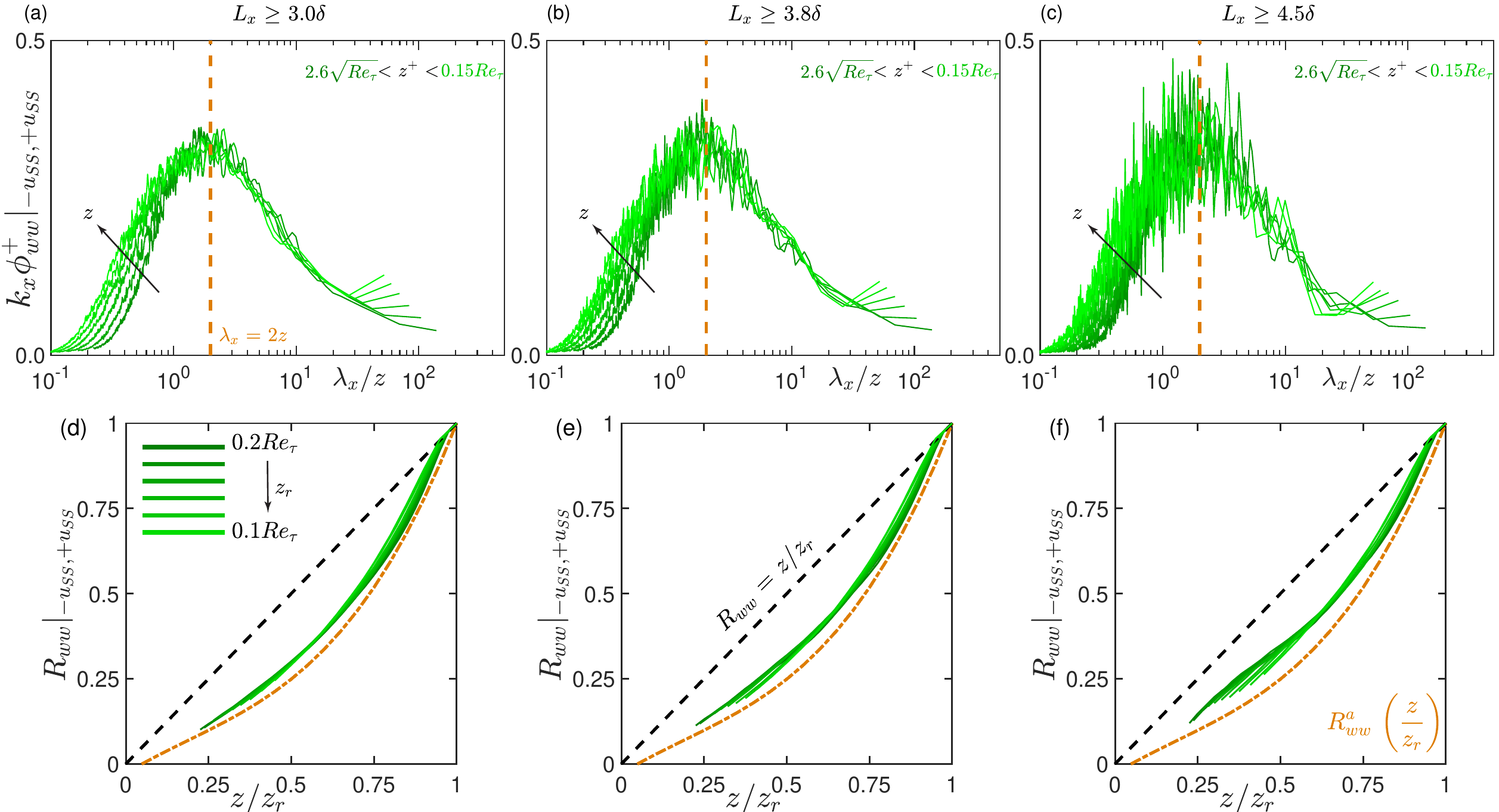}}
  \caption{(a-c) Conditionally averaged premultiplied 1-D spectra of $w$-fluctuations plotted vs ${{\lambda}_{x}}/{z}$ at various $z^{+}$ within the log-region. 
	(d-f) Correlation between $w$-fluctuations at $z$ and $z_{r}$, normalized by ${\overline{w^2}}({z_r})$ for various $z_{r}$.
  Here, both the spectra and the cross correlations are computed from the extracted $w$-flow fields associated with -${u_{ss}}$ and +${u_{ss}}$ detected in the LFOV PIV data at $Re_{\tau}$ $\approx$ 7500, for varying thresholds of the streamwise lengths associated with the superstructures: (a,d) $L_{x}$ $\gtrsim$ 3$\delta$, (b,e) $L_{x}$ $\gtrsim$ 3.8$\delta$ and (c,f) $L_{x}$ $\gtrsim$ 4.5$\delta$.
	In (a-c), the {\color{Golden}dashed golden} line represents the linear scaling, ${{\lambda}_{x}}$ $=$ 2${z}$.
	In (d-f), the dashed black line corresponds to the linear relationship, $z/{z_{r}}$ while the {\color{Golden}dash-dotted golden} line corresponds to $R^{a}_{ww}$ defined in (\ref{eq2}).}	
\label{fig15}
\end{figure}

In figure \ref{fig14}, the spectra are plotted for $z^+$ $\approx$ 2.6$\sqrt{Re_{\tau}}$ (nominal start of the log-region) and the middle of the boundary layer ($\approx$ 0.5${Re_{\tau}}$), and compared against published spectra estimated from multiwire experiments \citep{winter2015,baidya2017} conducted in the same facility and at similar $Re_{\tau}$.
It should be noted here that the hotwire spectra are plotted based on assumption of Taylor's hypothesis, with the mean velocity at $z^+$ considered as the mean convection velocity of the turbulent scales.
Taylor's approximation has been deemed a reasonable assumption in the inertial region for ${\lambda}_{x}$ $\lesssim$ 6$\delta$ \citep{dennis2008,charitha2015}, permitting its use for validation of the PIV spectra up to this scale range.
One can clearly observe that the PIV and hotwire spectra match reasonably well, especially in the intermediate-scale range ($\mathcal{O}(1)$ $\lesssim$ ${{\lambda}_{x}}/{z}$ $\lesssim$ $\mathcal{O}(10)$), which nominally corresponds to the self-similar hierarchy coexisting in the boundary layer.
Further, the agreement also confirms the insignificant influence of the evolution of the boundary layer thickness on the spectra, which is estimated by extracting data at a constant $z$-location along the $x$-direction.
Some discrepancy at smaller ${\lambda}_{x}$ is expected for high $Re_{\tau}$ PIV ($\sim$ 5000, 7500), due to relatively poor spatial resolution of PIV compared to the hotwire sensor resolution (table \ref{tab1}).

\section*{Appendix 2: Effect of thresholds on conditional statistics}
\label{app2}

Figures \ref{fig15}(a-c) present the conditionally averaged, premultiplied $w$-spectra and figures \ref{fig15}(d-f) present the conditionally averaged two-point correlations of the $w$-fluctuations.
Both are computed from the flow fields extracted based on varying thresholds ($L_{x}$) on the streamwise extents of the identified superstructures.
The statistics are computed for the $Re_{\tau}$ $\approx$ 7500 LFOV PIV dataset at $z^+$ corresponding to the log-region.
It is evident from the comparison that the $z$-scalings, exhibited by both the statistics, remain unchanged despite the change in $L_{x}$.
The most prominent effect of increasing the $L_{x}$ is the reduced number of ensembles (of the extracted flow fields), leading to poorly converged conditionally averaged statistics.

\bibliographystyle{jfm}
\bibliography{ConcatenationAE_formSS_bib}

\end{document}